\documentclass[iop,apj]{emulateapj}

\usepackage{aas_macros}
\pdfoutput=1 
\usepackage{amsmath,amstext}
\usepackage[T1]{fontenc}
\usepackage{apjfonts} 
\usepackage[normalem]{ulem}

\usepackage{color}

\DeclareMathAlphabet{\mathsc}{OT1}{cmr}{m}{sc}
\def\testbx{bx}%
\DeclareRobustCommand{\ion}[2]{%
\relax\ifmmode
\ifx\testbx\f@series
{\mathbf{#1\,\mathsc{#2}}}\else
{\mathrm{#1\,\mathsc{#2}}}\fi
\else\textup{#1\,{\mdseries\textsc{#2}}}%
\fi}

\shorttitle{$N_{\rm H}/E(B-V)$}
\shortauthors{Lenz, Hensley, and Dor\'e}

\begin{document}

\title{A new, large-scale map of interstellar reddening derived from \ion{H}{i} emission}

\author{Daniel Lenz}
\author{Brandon S. Hensley}
\author{Olivier Dor\'e}

\email{daniel.lenz@jpl.nasa.gov}
\email{brandon.s.hensley@jpl.nasa.gov\\
\copyright 2017 California Institute of Technology\\U.S. Government sponsorship acknowledged}
\affiliation{Jet Propulsion Laboratory, California Institute of Technology, 4800
Oak Grove Drive, Pasadena, CA 91109, USA}

\begin{abstract}
We present a new map of interstellar reddening, covering the 39\% of the sky with low \ion{H}{i} column densities ($N_\ion{H}{i} < 4\times10^{20}\,$cm$^{-2}$ or $E(B-V)\approx 45$\,mmag) at $16\overset{'}{.}1$ resolution, based on all-sky observations of Galactic \ion{H}{i} emission by the HI4PI Survey. In this low column density regime, we derive a characteristic value of $N_\ion{H}{i}/E(B-V) = 8.8\times10^{21}\,\rm\,cm^{2}\,mag^{-1}$ for gas with $|v_{\rm LSR}| < 90$\,km\,s$^{-1}$ and find no significant reddening associated with gas at higher velocities. We compare our \ion{H}{i}-based reddening map with the Schlegel, Finkbeiner, and Davis (1998, SFD) reddening map and find them consistent to within a scatter of $\simeq 5$\,mmag. Further, the differences between our map and the SFD map are in excellent agreement with the low resolution ($4\overset{\circ}{.}5$) corrections to the SFD map derived by Peek and Graves (2010) based on observed reddening toward passive galaxies. We therefore argue that our \ion{H}{i}-based map provides the most accurate interstellar reddening estimates in the low column density regime to date.
Our reddening map is made publicly available\footnote{\url{http://dx.doi.org/10.7910/DVN/AFJNWJ}}. 
\end{abstract}


\section{Introduction}
Gas and dust are well-mixed in the Galactic interstellar medium (ISM), as evidenced by the tight empirical correlation between 21\,cm \ion{H}{i} emission and both dust emission \citep[e.g.,][]{Low+etal_1984, Boulanger+etal_1996} and extinction \citep[e.g.,][]{Sturch_1969, Bohlin+Savage+Drake_1978}. Consequently, measurements of optical dust reddening are frequently used to infer gas column densities toward sources of interest. Likewise, tracers of dust column, such as far-infrared (FIR) dust emission or \ion{H}{i} emission, are converted to maps of optical reddening which are widely used to correct extragalactic observations for Galactic extinction.

Optical reddening is generally expressed as the difference in observed extinction between the Johnson B and V bands, i.e., $E(B-V) \equiv A_B - A_V$. As dust is more effective at absorbing and scattering light in the B versus V band (i.e., $A_B > A_V$), $E(B-V)$ increases with increasing dust column. To the extent that the B and V band dust opacities do not vary spatially, this correlation will be linear. Direct measurement of $E(B-V)$ typically involves measuring emission from a source of known spectrum (e.g. quiescent galaxies, stars) and attributing deviations from the known spectrum to dust attenuation.

Producing {\it maps} of interstellar reddening in this way is challenging. Many observations, ideally spectra, of sources of known spectral energy distributions (SEDs) are required both to cover the requisite area with sufficient density and also to minimize errors due to intrinsic spectral variations in the sources. Further, if the sources employed are Galactic, care must be taken to ensure that all are behind the full Galactic dust column or else the total reddening will be underestimated. 

The most extensive and sensitive large scale map of optical reddening to date is based on Pan-STARRS1 photometry toward more than 500 million stars. Employing the techniques of \citet{Green+etal_2014} to infer distance and reddening to stars based on the photometry, \citet[][hereafter S14]{Schlafly+etal_2014} used the Pan-STARRS1 data to construct a map of interstellar reddening to a distance of 4.5\,kpc. The typical uncertainties of this map are $\simeq 20$\,mmag, arising both from the stellar photometry and the number of stellar sources available. Subsequently, a full three-dimensional map of dust in the Milky Way has also been derived from these data \citep{Green+etal_2015}.

Because of these intrinsic limitations, the most widely used maps of optical reddening are generally created via a different approach. Since $E(B-V)$ is proportional to the dust column, direct measurements of other observables proportional the dust column can be converted to $E(B-V)$ if the conversion factor is known. A small set of objects with directly measured $E(B-V)$ can be used to calibrate a well-measured tracer of the dust column, such as (FIR) dust emission or \ion{H}{i} emission, along the same lines of sight. This relation can then be extrapolated to all lines of sight along which the dust column tracer has been measured. At present, the signal-to-noise ratio of observations of FIR dust emission and \ion{H}{i} emission is superior to direct measurements of $E(B-V)$ from stellar data, and therefore the resulting reddening maps typically have much smaller observational uncertainty. The limiting factor of this technique is the calibration to $E(B-V)$.

Beginning with the seminal work of \citet[][hereafter SFD]{Schlegel+Finkbeiner+Davis_1998}, FIR dust emission has been the preferred tracer of the dust column from which the optical extinction is inferred. A primary advantage of this approach is the availability of sensitive, high-resolution, full-sky maps of dust emission. SFD employed both IRAS and DIRBE data along with measured optical extinction toward a sample of elliptical galaxies to calibrate a relation between FIR flux and $E(B-V)$. Most recently, \citet{Planck_2013_XI} and \citet[][hereafter MF]{Meisner+Finkbeiner_2015} employed high-sensitivity measurements of dust emission from the {\it Planck} satellite to derive full-sky maps of the dust optical depth and, in turn, $E(B-V)$.

Comparing the reddening predicted by the SFD map to direct photometric reddening measurements of quiescent galaxies, \citet{Peek+Graves_2010} found typical discrepancies of only $\pm3$\,mmag at a resolution of $4\overset{\circ}{.}5$. However, they also noted a few extended regions having residuals of up to 45\,mmag that coincided with low dust temperatures. In regions of the SFD map with $E(B-V) \lesssim 0.1$, \citet{Yahata+etal_2007} noted that galaxy number counts {\it decreased} with decreasing reddening, rather than increasing. They also found that galaxies became slightly {\it bluer} with increasing SFD reddening in these regions. They therefore suggested that the SFD reddenings are contaminated at a low level ($\sim 10$\,mmag, corresponding to tens of percent in these regions) by extragalactic infrared emission being attributed to Galactic dust. In higher extinction regions near the Galactic plane, systematic discrepancies between SFD and stellar reddenings have also been observed \citep{Schlafly+etal_2014}.

These discrepancies are not unexpected as the connection between infrared emission and optical extinction is not completely straightforward. First, the dust emission at a given frequency depends not only upon the dust column but also on the dust temperature. Determining the dust temperature requires modeling of the full dust SED, particularly near the peak where temperature effects are strongest. The temperature determination is further complicated by the fact that dust emission along the line of sight may encompass several interstellar environments each with their own characteristic dust temperature. Additionally, dust species of different compositions and different sizes likely achieve different equilibrium temperatures. Thus, correcting for temperature effects can be challenging.

Second, the Galactic FIR dust emission is contaminated both by the Cosmic Infrared Background (CIB) and the zodiacal light. Inaccurate subtraction of either of these components will induce errors in the inferred dust temperatures and/or column densities. Indeed, MF identified these emission components as the limiting factors in deriving reddening maps based on the {\it Planck} data.

Finally, even with perfect temperature corrections and CIB and zodiacal light subtraction, the amount of dust reddening per unit dust column is a variable quantity within the Galaxy as attested by systematic variations in $R_V \equiv A_V/E(B-V)$ \citep{Cardelli+Clayton+Mathis_1989, Fitzpatrick+Massa_2007, Schlafly+etal_2016}. Any method based solely upon the total dust column will be unable to account for these variations.

These limitations motivate us to revisit \ion{H}{i} emission as a predictor of interstellar reddening. The use of \ion{H}{i} in this context was pioneered by \citet{Burstein+Heiles_1978} and \citet{Burstein+Heiles_1982}. Since \ion{H}{i} cannot trace variations in the gas to dust ratio, they supplemented the \ion{H}{i} observations with galaxy counts to estimate reddening, as the number of observed galaxies per square degree decreases as reddening increases. In this way they produced a large scale map of interstellar reddening with quoted accuracy as low as 10\,mmag, though SFD noted a global 20\,mmag zero point offset relative to the FIR-based reddening maps. \citet{Burstein+Heiles_1978} identified large, coherent variations in the dust to \ion{H}{i} ratio throughout the Galaxy. They argued that much of this variation could be accounted for by the presence of H$_2$, though sightlines with anomalous dust to gas ratios were noted even at low column densities. Such variations could prove a fundamental limitation to reddening determinations based solely upon \ion{H}{i}. Indeed, combining \ion{H}{i} data from the LAB Survey \citep{Kalberla+etal_2005} and the SFD reddening map, \citet{Peek_2013} argued that a combination of \ion{H}{i} and dust emission data provided a more accurate predictor of reddening than either tracer alone.

The availability of new, high-sensitivity, full-sky maps of \ion{H}{i} emission from the HI4PI Survey \citep{Hi4pi_2016} enables a reinvestigation of the \ion{H}{i}-$E(B-V)$ relationship in unprecedented detail. The use of the gas column allows us to circumvent several issues associated with methods based on FIR dust emission. First, the gas column is measured directly without the need for dust temperature corrections. Second, emission from the CIB or zodiacal light is not a contaminant for the \ion{H}{i} data. Finally, contamination from extragalactic \ion{H}{i} sources can be largely removed with simple velocity cuts. 

The use of \ion{H}{i} data as a tracer of reddening is, however, limited by the extent to which $N_\ion{H}{i}$ is correlated with the dust column. The presence of H$_2$ (and consequently H$_2$-correlated dust) along the line of sight and variations in the dust to gas ratio are both problematic for this method. However, the former can be largely circumvented by working at low column densities. The latter, as we demonstrate in this work, can be mitigated through use of the \ion{H}{i} velocity information to account for systematic differences that exist in reddening per H atom between moderate and high velocity gas \citep[e.g.][]{Planck_2011_XXIV}.

At low column densities ($N_\ion{H}{i} < 4\times10^{20}\,$cm$^{-2}$), where \ion{H}{i} composes most of the gas column, the reddening map derived in this work is in excellent agreement with the dust-based reddening maps to a level of $\simeq 5$\,mmag. Our residuals with respect to the SFD map agree well with the corrections to SFD derived by \citet{Peek+Graves_2010} based on extragalactic reddening measurements. Further, the SFD reddenings are known to be correlated with large scale structure \citep{Yahata+etal_2007}, while the \ion{H}{i}-based reddenings do not suffer from this systematic. Thus, we advocate the use of our \ion{H}{i}-based reddening map in the 39\% of the sky with $N_\ion{H}{i} < 4\times10^{20}\,$cm$^{-2}$.

This paper is organized as follows: in Section~\ref{sec:data}, we describe the datasets used in this analysis; in Section~\ref{sec:hi_ebv}, we explore the relationship between $N_\ion{H}{i}$ and $E(B-V)$ as a function of column density to inform our data model; in Section~\ref{sec:reddening_map} we quantify the velocity dependence of $N_\ion{H}{i}$/$E(B-V)$ and construct a reddening map from the \ion{H}{i} data, which we compare to existing reddening maps calibrated on far-infrared dust emission in Section~\ref{sec:compare}; we discuss potential extensions of this work in Section~\ref{sec:discussion}; and in Section~\ref{sec:conclusions}, we summarize our principal conclusions.

\section{Data}
\label{sec:data}

In the following, we give a brief introduction of the different data sets that are used in this work. We perform all analysis on HEALPix\footnote{\url{http://healpix.sf.net/}} maps \citep{Gorski+etal_2005} with $\rm N_{side}=512$, corresponding to an approximate pixel size of $6\overset{'}{.}9$. We smooth all data sets with a Gaussian of FWHM $16\overset{'}{.}1$, corresponding to the resolution of the \ion{H}{i} data. Lower resolution data products are left at their native resolution.

\subsection{$E(B-V)$}
\subsubsection{Schlegel, Finkbeiner, and Davis 1998 (SFD)}

The SFD reddening map\footnote{\url{https://lambda.gsfc.nasa.gov/product/foreground/fg\_sfd\_get.cfm}}, based on the FIR observations of \textit{COBE}/DIRBE and IRAS \citep{Silverberg+etal_1993, Neugebauer+etal_1984}, has been the state-of-the-art map of stellar reddening since its release, covering the full sky at an angular resolution of $6\overset{'}{.}1$. To recover the actual dust column density based on the 100\,$\mu$m IRAS data, a temperature correction based on DIRBE is applied on angular scales of 1$^\circ$, corresponding to the DIRBE angular resolution. Further corrections are made for Zodiacal light, the CIB, and the scanning pattern. The reddening is derived by calibrating the $100\,\rm\mu m$ flux density against the colors of elliptical galaxies \citep{Postman+Lauer_1995}.

The SFD calibration was revisited by \citet{Schlafly+Finkbeiner_2011}, who, in agreement with the earlier results of \citet{Schlafly+etal_2010}, determined that the SFD $E(B-V)$ values were biased high. Subsequent reddening determinations with SDSS quasars also support the reddening vector of \citet{Schlafly+Finkbeiner_2011} over that used by SFD \citep{Wolf_2014}. Using the correction factors for the Landolt B and V bands in Table~6 of \citet{Schlafly+Finkbeiner_2011}, $E(B-V)_{\rm true} = 0.884\cdot E(B-V)_{\rm SFD}$. For all analysis presented in this work, we apply this correction to the SFD data as recommended by the current SFD release \citep{SFD_data} \footnote{\url{https://dataverse.harvard.edu/\\ dataset.xhtml?persistentId=doi:10.7910/DVN/EWCNL5}}.

\subsubsection{Schlafly et al 2014 (S14)}

The S14 reddening map\footnote{\url{https://lambda.gsfc.nasa.gov/product/foreground/fg\_ebv\_map\_info.cfm}} is based on Pan-STARRS1 optical photometry of 500 million stars, giving the integrated reddening to 4.5\,kpc for the entire northern sky ($\delta > -30^{\circ}$). While this distance limit underestimates the full dust column in the Galactic disk, it is sufficient to fully trace the dust and intermediate and high Galactic latitudes. The model applied to the Pan-STARRS1 photometry to simultaneously estimate the reddening and the distance to each star is described in \citet{Green+etal_2014}. \citet{Schlafly+etal_2014} find good overall agreement with the SFD map and estimate the reddening uncertainty of their map to be $\simeq$25\,mmag. At high Galactic latitudes ($|b| \gtrsim 30^\circ$), the noise is increased by the low stellar density. Therefore, at the low column densities central to this work, the reddening uncertainties in S14 are much larger than those in maps based on the FIR dust emission.

The resolution and the sensitivity of the S14 map depend on the number of stars for each resolution element for which the reddening is modeled. Due to the greater number of stars closer to the Galactic disk, the angular resolution varies from $7'$ ($\rm N_{side}=512$) at low and intermediate Galactic latitudes to $14'$ ($\rm N_{side}=256$) at the Galactic poles. For the work presented here, we re-sample this data set to a homogeneous grid with $\rm N_{side}=512$. As this map is on the same reddening scale as SFD, we likewise employ a correction of $E(B-V)_{\rm true} = 0.884\cdot E(B-V)_{\rm S14}$ for all analyses.

\subsubsection{Meisner \& Finkbeiner 2015 (MF)}
\label{subsubsec:MF}
The MF reddening map\footnote{\url{https://faun.rc.fas.harvard.edu/ameisner/planckdust/}} is derived from a dust model consisting of two modified blackbodies of different temperatures and spectral indices. This model is used to perform a pixel-by-pixel, full-sky fit of the FIR data from \textit{Planck} 2013 PR1 and DIRBE/IRAS \citep{Neugebauer+etal_1984, Silverberg+etal_1993, Planck_2011_VI}, resulting in full-sky maps of dust optical depth and dust temperature at an angular resolution of $6\overset{'}{.}1$. Their map of 545\,GHz dust optical depth $\tau_{545}$ is converted to $E(B-V)$ by calibrating against reddening inferred from stellar spectroscopy and photometry. For the stellar data, they employ roughly 260,000 stars with spectroscopy and broadband photometry from the SDSS's SEGUE Stellar Parameter Pipeline (SSPP) with reddenings determined by \citet{Schlafly+Finkbeiner_2011}, resulting in a final relation of $E(B-V)\,[\rm mag] = 2.62\times10^3\tau_{545}$.

For our analysis, we smooth this map to $16\overset{'}{.}1$ resolution with a Gaussian kernel and downgrade it to $\rm N_{side}=512$, using a simple average in each pixel. As this map is on the same reddening scale as SFD, we employ a correction of $E(B-V)_{\rm true} = 0.884\cdot E(B-V)_{\rm MF}$ for all analyses.

\subsubsection{Other Reddening Maps}
\citet{Peek+Graves_2010} employed SDSS photometry of $\sim 150,000$ passive galaxies with standard intrinsic colors to construct a map of deviations from the SFD map. They found a typical difference of around 3\,mmag at a resolution of $4\overset{\circ}{.}5$, though as high as 45\,mmag in the most problematic regions. As the discrepancies appeared correlated with the SFD temperature correction, the residuals are likely due to imperfections in the SFD map. In this study, we employ the \citet{Peek+Graves_2010} reddening map as a benchmark to assess the \ion{H}{i}-based reddenings derived in this work. 

We use the SFD corrections by \citet{Peek+Graves_2010} in zenith equal-area projection\footnote{Peek, priv. comm.}, which we reproject to HEALPix with $\rm N_{side}=512$ using the \texttt{reproject}\footnote{\url{https://reproject.readthedocs.io}} package. We leave the map at its native $4\overset{\circ}{.}5$ resolution. As these corrections were performed on the same reddening scale as SFD, we apply the correction factor of 0.884 to this map for all analysis.

Finally, \citet{Planck_2013_XI} constructed full-sky maps of interstellar reddening by performing dust model fits to the {\it Planck} and IRAS data, then correlating the model parameters with reddenings measured toward 53,399 SDSS quasars. As the 353\,GHz optical depth $\tau_{353}$ was found to be affected by CIB anisotropies at low column densities, the dust radiance $\mathcal{R} \equiv \int I_{\nu}\,\mathrm{d}\nu$ was preferred as an estimate of the dust column. Comparing the {\it Planck} $E(B-V)$ map to stellar reddening data, both S14 and MF find large scale systematic residuals. We provide a further test of this map in Section~\ref{sec:compare}.

\subsection{\ion{H}{i}}

For \ion{H}{i} emission, we employ data from the recently released full-sky spectroscopic HI4PI Survey \citep{Hi4pi_2016}\footnote{\url{http://cdsarc.u-strasbg.fr/viz-bin/qcat?J/A+A/594/A116}}, which merges the data from the Effelsberg-Bonn \ion{H}{i} Survey \citep[EBHIS,][]{Winkel+etal_2010, Kerp+etal_2011, Winkel+etal_2016} and the Galactic All-Sky Survey \citep[GASS,][]{Mcclure-Griffiths+etal_2009, Kalberla+etal_2010, Kalberla+Haud_2015}. This dataset covers emission from neutral atomic hydrogen over the full sky with radial velocities $|v_{\rm LSR}| < 600\,\rm km\,s^{-1}$ ($470\,\rm km\,s^{-1}$ for the southern hemisphere, covered by GASS) with a spectral resolution of $1.45\,\rm km\,s^{-1}$.

This full-sky survey supersedes the Leiden/Argentine/Bonn (LAB) Survey of Galactic \ion{H}{i} \citep{Kalberla+etal_2005} and offers a superior angular resolution ($16.1'$ versus $36'$), higher sensitivity (43 versus $80\,\rm mK$), and better spatial sampling (full angular sampling vs. beam-by-beam sampling). Moreover, the corrections for stray radiation have been shown to be reliable even in low-column density regions \citep{Martin+etal_2015}, where the uncorrected stray radiation is often stronger than the actual signal \citep[][Figures~1 and A.2]{Winkel+etal_2016}.

For our analysis, we downgrade the HEALPix resolution of the HI4PI data from $\rm N_{side}=1024$ to 512.

\section{The $E(B-V)$/$N_\ion{H}{i}$ ratio}
\label{sec:hi_ebv}

\begin{figure*}[tp]
	\includegraphics[width=\columnwidth]{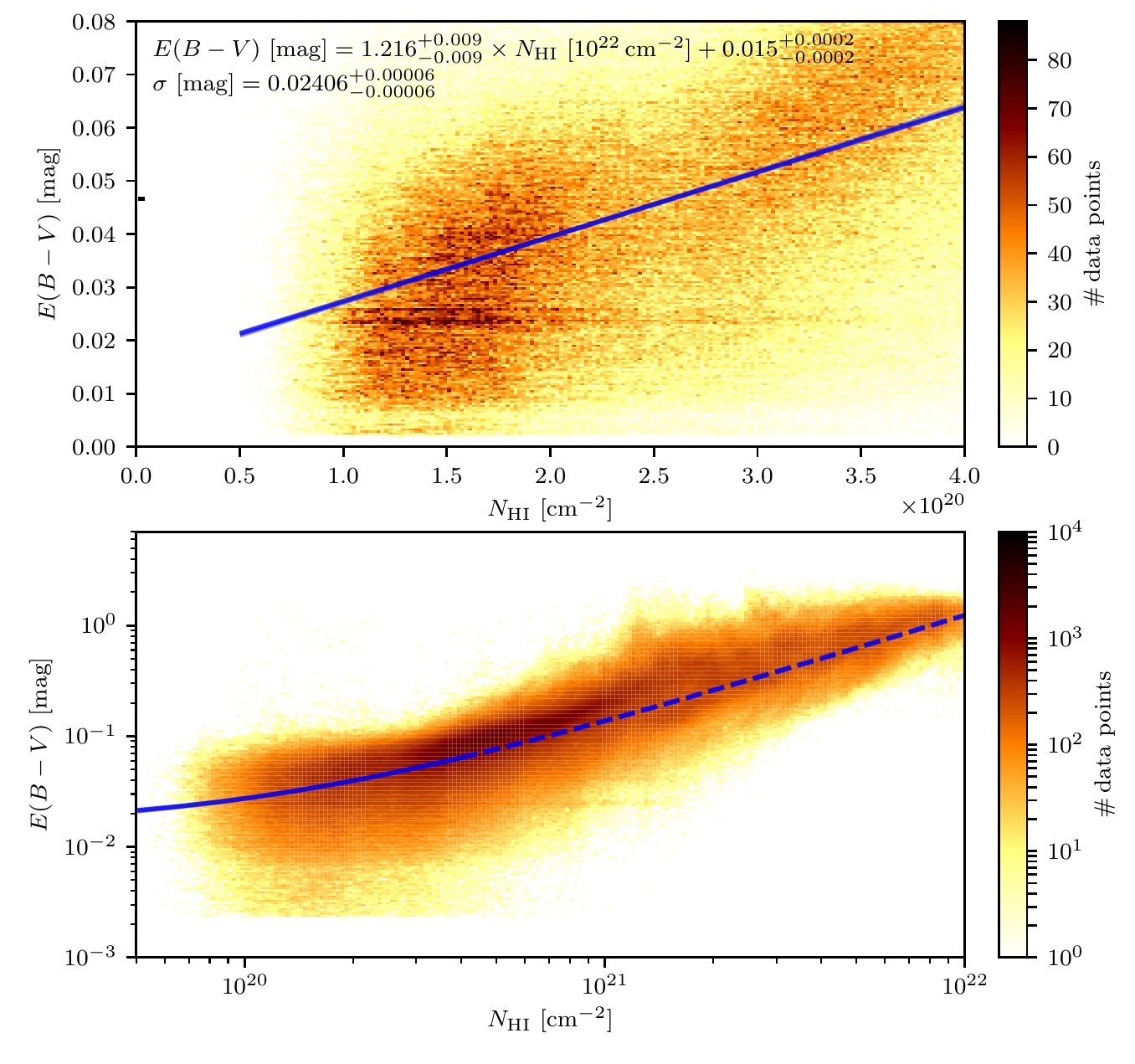}
    \includegraphics[width=\columnwidth]{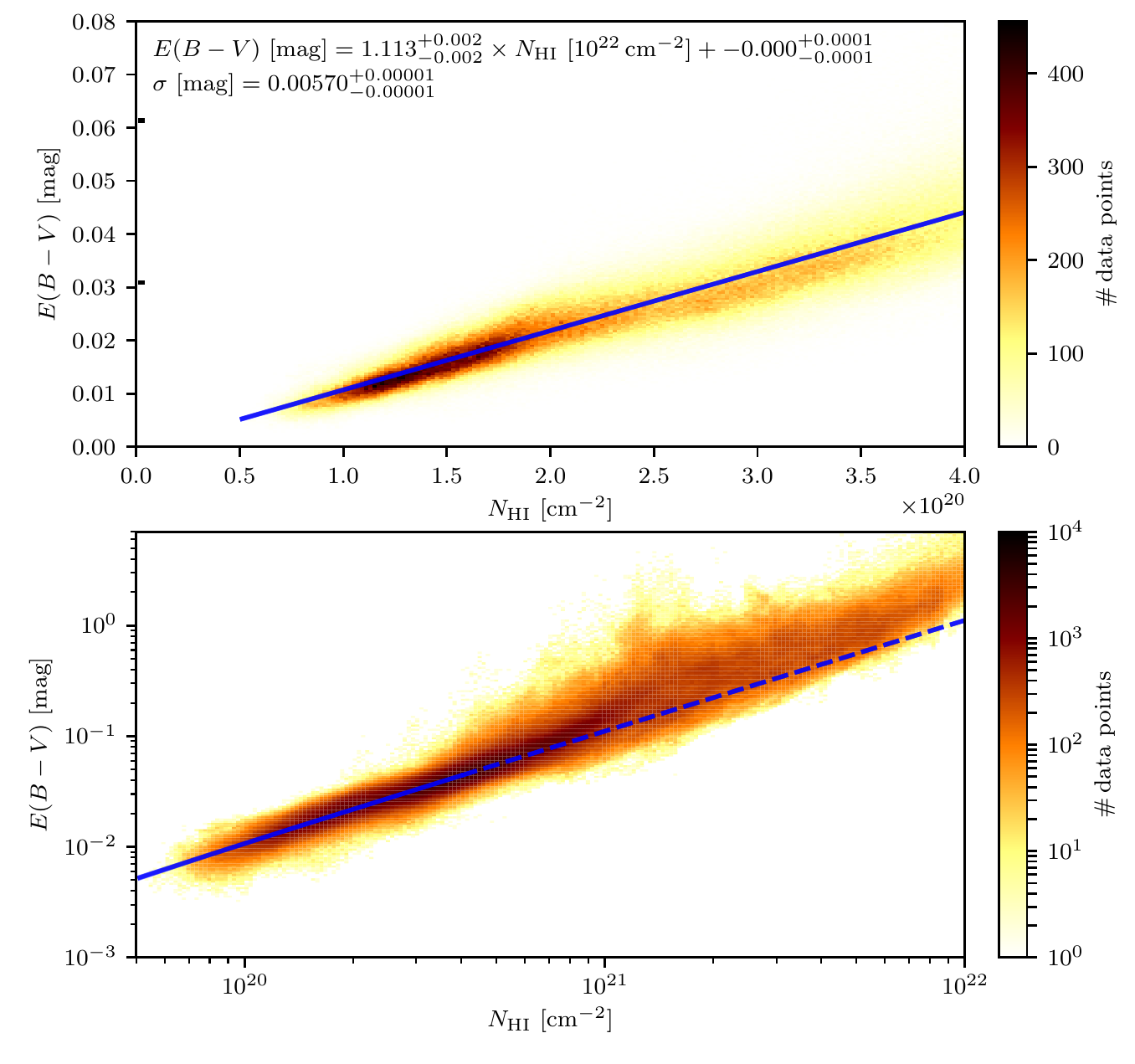}
	\caption{We plot the total \ion{H}{i} column density in each pixel against the reddening in that pixel as measured from stellar data by S14 \textbf{left} and from FIR dust emission by SFD \textbf{right}. Note that the S14 comparison is limited to the S14 footprint whereas the SFD comparison covers the full sky. The top panels show the data on a linear scale and restricted to $N_\ion{H}{I} < 4\times10^{20}$\,cm$^{-2}$, while the bottom panels are logarithmically scaled and demonstrate the deviation from linearity at high $N_\ion{H}{i}$. The blue line indicates the fit to the data for $N_\ion{H}{i} < 4\times10^{20}$\,cm$^{-2}$ (solid) and is extrapolated to higher column densities (dashed).}
	\label{fig:ebv_vs_nhi}
\end{figure*}

It has long been appreciated that there is a roughly linear correlation between the \ion{H}{i} column density $N_\ion{H}{i}$ and the reddening $E(B-V)$. This scaling is not unexpected -- $A_\lambda$ is proportional to the dust column at fixed dust opacity $\kappa_\lambda$, and, insofar as dust and gas are well-mixed in the ISM, $N_\ion{H}{i}$ also scales linearly with the dust column. This scaling breaks down at high $N_\ion{H}{i}$ where an appreciable fraction of the gas is molecular \citep{Wolfire+etal_2010}. The measured value of $E(B-V)$/$N_\ion{H}{i}$ is widely used in both Galactic and extragalactic astronomy to convert measured optical reddening to an estimate of the gas column.

In this Section, we explore the relationship between $N_\ion{H}{i}$ and $E(B-V)$ in the context of existing reddening maps, focusing in particular on the linearity of the relationship and its breakdown at high column densities.

\subsection{Previous Determinations}
Early determinations of $N_\ion{H}{i}/E(B-V)$ were made by correlating 21\,cm emission with observed reddening in various sources (e.g. RR Lyrae, globular clusters, A stars), arriving at a characteristic value of $\sim5\times10^{21}$\,cm$^{-2}$\,mag$^{-1}$ \citep{Sturch_1969,Knapp+Kerr+1974,Mirabel+Gergely_1979,Heiles_1976}. Several of these studies required a statistically significant nonzero intercept in the relation, hinting at different behavior at low reddening. 

Ly$\alpha$ absorption measurements have allowed complementary determinations of the \ion{H}{i} column and thus independent determinations of the relationship between the hydrogen column and $E(B-V)$. \citet{Jenkins+Savage_1974} found $N_\ion{H}{i}/E(B-V) = 5.3\times10^{21}$\,cm$^{-2}$\,mag$^{-1}$ via this method and $N_{\rm H, total}$ = $N_{\ion{H}{i}}$ + $2N_{{\rm H}_2}$ + $N_\ion{H}{ii} = 7.5\times10^{21}E(B-V)$\,cm$^{-2}$\,mag$^{-1}$ when restricting the analysis to sightlines where H$_2$ formation was unimportant and applying a correction to account for \ion{H}{ii}. With the latter relation, they are able to infer $N_\ion{H}{i}$ given $E(B-V)$ to an rms accuracy of $2\times10^{20}$\,cm$^{-2}$, corresponding to 27\,mmag of reddening.

Employing Ly$\alpha$ absorption measurements toward 75 early type stars, \citet{Bohlin+Savage+Drake_1978} derived $N_{\ion{H}{i}+{\rm H}_2}/E(B-V) = 5.8\times10^{21}$\,cm$^{-2}$\,mag$^{-1}$ with an estimated 6\% uncertainty, a value consistent with previous determinations and which is frequently quoted in the literature.

More recent studies employing Ly$\alpha$ measurements from the {\it International Ultraviolet Explorer} yielded $N_\ion{H}{i}/E(B-V) \simeq 5\times10^{21}$\,cm$^{-2}$\,mag$^{-1}$  \citep{Shull+vanSteenberg_1985, Diplas+Savage_1994}, while determinations of the H$_2$ column density with the {\it Far Ultraviolet Spectroscopic Explorer} were employed to derive $N_{\ion{H}{i}+{\rm H}_2}/E(B-V) = \left(5.94\pm0.37\right)\times10^{21}$\,cm$^{-2}$\,mag$^{-1}$, in good agreement with the results of \citet{Bohlin+Savage+Drake_1978}.

However, this consensus was recently challenged by \citet{Liszt_2014a,Liszt_2014b} who correlated the \ion{H}{i} emission measured by the Leiden/Argentine/Bonn Survey \citep{Kalberla+etal_2005} with the SFD reddening map and found a characteristic value of $N_\ion{H}{i}/E(B-V) = 
8.3\times10^{21}$\,cm$^{-2}$\,mag$^{-1}$ for $|b| \geq 20^\circ$. \citet{Liszt_2014a} notes good agreement between this value and low-reddening estimates of previous studies. For instance, when employing the linear fit derived by \citet{Heiles_1976} at $E(B-V) \lesssim 0.1$\,mag, larger $N_\ion{H}{i}/E(B-V)$ values consistent with $8.3\times10^{21}$\,cm$^{-2}$\,mag$^{-1}$ are obtained. Likewise, the Ly$\alpha$ measurements of \citet{Diplas+Savage_1994} mostly sampled sightlines with $E(B-V) > 0.1$\,mag, with the lower reddening sightlines showing some evidence of different behavior. However, the sightlines with $E(B-V) < 0.1$\,mag observed by \citet{Bohlin+Savage+Drake_1978} fell on the same $N_\ion{H}{i}-E(B-V)$ relation as those sightlines at higher reddening.

Equipped with new and superior full-sky \ion{H}{i} observations and extensive maps of direct and indirect measurements of interstellar reddening, we revisit the $N_\ion{H}{i}$-$E(B-V)$ relationship in detail with particular emphasis on how the slope of the relation changes with increasing $E(B-V)$.

\subsection{A New Determination}

In the left panels of Figure~\ref{fig:ebv_vs_nhi}, we plot $N_\ion{H}{i}$ from HI4PI and $E(B-V)$ as determined by S14 for every pixel in the S14 footprint. While there is large scatter in the measured optical reddenings, a linear relationship is evident. Fitting only those pixels with $N_\ion{H}{i} < 4\times10^{20}$\,cm$^{-2}$, we find a slope of $E(B-V)/N_\ion{H}{i} = \left(1.22\pm0.01\right)\times10^{-22}$\,cm$^2$\,mag, or equivalently $N_\ion{H}{i}/E(B-V) = 8.2\times10^{21}$\,cm$^{-2}$\,mag$^{-1}$. This value is in much closer agreement with that of \citet{Liszt_2014a} than \citet{Bohlin+Savage+Drake_1978} (8.3 and 5.8$\times10^{21}$\,cm$^{-2}$\,mag$^{-1}$, respectively). We fit an intercept of 15\,mmag and an intrinsic scatter of 24\,mmag, both of which are consistent with the quoted $\sim25$\,mmag uncertainty on the optical reddening values \citep{Schlafly+etal_2014}. Indeed, when comparing their measured reddenings to the SFD map, \citet{Schlafly+etal_2014} subtracted an offset of 30\,mmag from their reddening map.

\begin{figure}[tp]
	\includegraphics[width=\columnwidth]{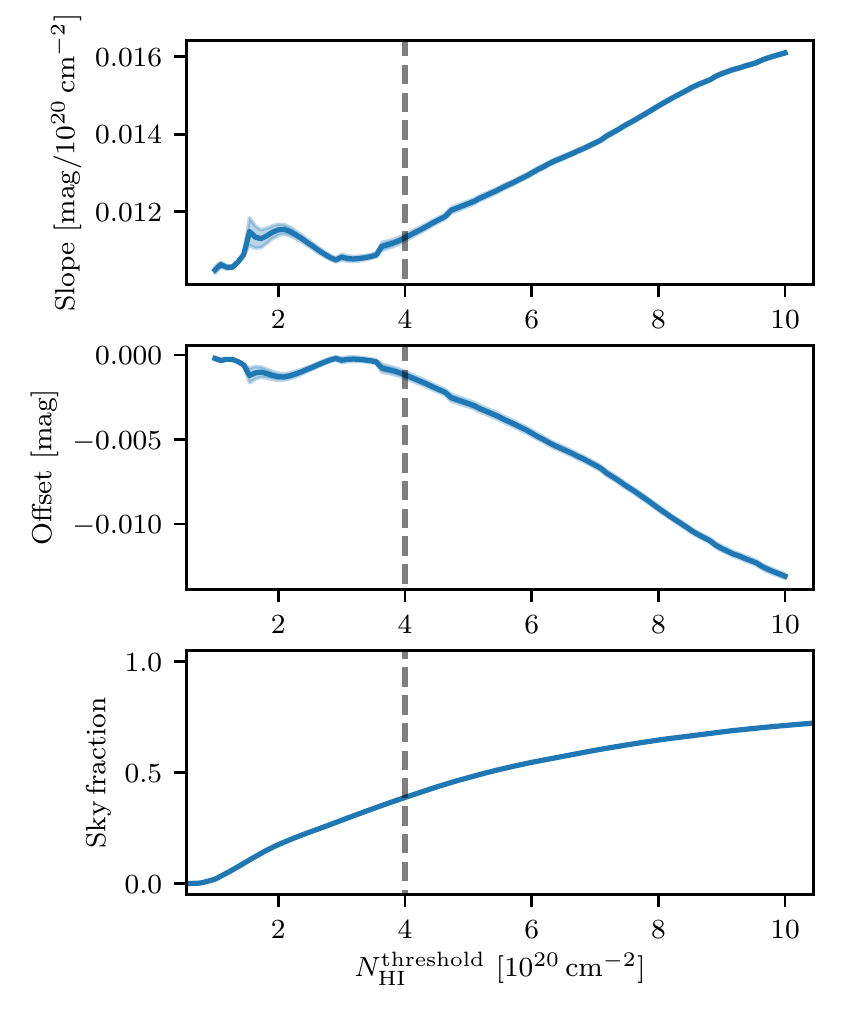}
	\caption{Fit parameters for the linear relation between $N_\ion{H}{i}$ and $E(B-V)$ as determined by SFD as a function of the $N_{\ion{H}{i}}$ threshold up to which the data are fit. The fit is relatively stable for  $N^{\rm threshold}_\ion{H}{I} < 4\times10^{20}$\,cm$^{-2}$, with the slope (\textbf{top}) varying by roughly 10\% in this range and the offset (\textbf{middle}) by only $\sim1-2$\,mmag. However, the slope and offset become degenerate above that value indicating departure from linearity. We furthermore show the fraction of the full sky for a given $N_{\ion{H}{i}}$ threshold (\textbf{bottom}). The vertical dashed line indicates our choice of the threshold for the subsequent analysis.}
	\label{fig:ebv_params}
\end{figure}

In the right panels of Figure~\ref{fig:ebv_vs_nhi}, we perform the same fit using the full-sky SFD reddening map instead of S14. At these low column densities (and thus low reddenings), the relative uncertainties in the S14 stellar reddening measurements are much higher than in SFD due to both photometric uncertainties and the low stellar densities on these generally high latitude sightlines. In the regime $N_\ion{H}{i} < 4\times10^{20}$\,cm$^{-2}$, the $E(B-V)$-$N_\ion{H}{i}$ relationship is remarkably linear with a best fit slope of $\left(1.113\pm0.002\right)\times10^{-22}$\,cm$^2$\,mag, or equivalently $N_\ion{H}{i}/E(B-V) = 8.8\times10^{21}$\,cm$^{-2}$\,mag$^{-1}$, and intercept consistent with zero. The fit indicates only 5.7\,mmag of intrinsic scatter, suggesting a very tight coupling between gas and dust at low column densities. A similar analysis of the MF data yields fit parameters of $\left(1.11\pm0.07\right)\times10^{-22}$\,cm$^2$\,mag for the slope, $(0\pm 2)\rm \, mmag$ for the offset, and 18\,mmag for the scatter. The values obtained with the SFD and MF maps are somewhat higher than the $8.3\times10^{21}$\,cm$^{-2}$\,mag$^{-1}$ obtained by \citet{Liszt_2014a}, which can be understood largely by the 12\% correction to the SFD calibration derived by \citet{Schlafly+Finkbeiner_2011} which was not applied by \citet{Liszt_2014a}.

From the log-log plots in the lower panels of Figure~\ref{fig:ebv_vs_nhi}, it is evident that the fit linear relation underestimates the observed reddening at column densities above $\simeq 4\times10^{20}$\,cm$^{-2}$. This is expected -- higher column density gas has a higher molecular fraction and thus more dust correlated with molecular rather than atomic hydrogen. This value is consistent with the thresholds of $2 - 5\times10^{20}$\,cm$^{-2}$ found in the literature \citep[e.g.][]{Boulanger+etal_1996, Wolfire+etal_2010}.

Given the excellent agreement of the zero point between the \ion{H}{i} data and the SFD reddening map, as well as a consistent 15-30\,mmag offset found between the S14 reddening map and the SFD map, the MF map, and the \ion{H}{i} data, we argue that this offset is inherent to the S14 reddenings. As noted by S14, this may be due to their fitting requirement that $E(B-V)$ be positive, which can lead to bias at low $E(B-V)$.

In Figure~\ref{fig:ebv_params}, we explore the effects of using different $N_\ion{H}{i}$ thresholds when performing the fit to the SFD reddenings. Fitting only pixels with $N_\ion{H}{i} < 4\times10^{20}$\,cm$^{-2}$, we find variations in the fit slope of $\lesssim10\%$ and fit intercepts of less than 2\,mmag. However, as the threshold is increased beyond $4\times10^{20}$\,cm$^{-2}$, the fit slope systematically increases while the intercept systematically decreases. It is therefore possible to arrive at {\it any} $N_\ion{H}{i}$/$E(B-V)$ value between those of \citet{Bohlin+Savage+Drake_1978} and \citet{Liszt_2014a} depending simply on the range of column densities employed in the fit. Given the small variations of the fit results for $N_\ion{H}{i} < 4\times10^{20}$\,cm$^{-2}$, we recommend $N_\ion{H}{i}/E(B-V)$ = 8.8$\times10^{21}$\,cm$^{-2}$\,mag$^{-1}$ as a representative value for the diffuse ISM where the hydrogen is predominantly atomic with a systematic uncertainty of roughly 10\%. We note that our determinations using the SFD, MF, and S14 maps are consistent within this uncertainty.

\subsection{Full Sky Minimum $E(B-V)$}
Atomic hydrogen is ubiquitous in the Galaxy. There is not a single sight line without HI detection in HI4PI, with the lowest value of $N_\ion{H}{i} \simeq 5\times10^{19}$\,cm$^{-2}$ at an angular resolution of $16'.1$. Using the linear relation derived in Figure~\ref{fig:ebv_vs_nhi}, this corresponds to a full sky minimum of reddening of 5\,mmag. This agrees well with both the SFD and the MF reddening maps, which have very few pixels (less than 0.1\%) with $E(B-V) < 5$\,mmag. We note that at higher angular resolutions, lower column density regions are present \citep[e.g. the Lockman Hole,][]{Lockman+etal_1986}.

\subsection{$\tau_{545}/N_\ion{H}{i}$}
\begin{figure}[tp]
	\includegraphics[width=\columnwidth]{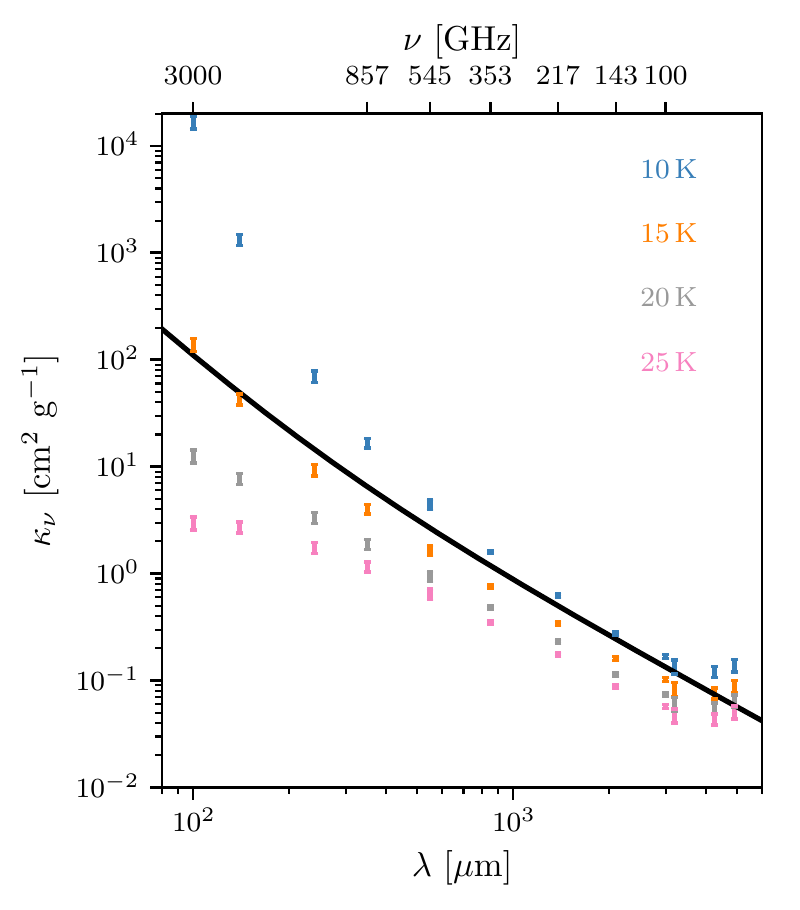}
	\caption{We plot the wavelength-dependent dust opacity implied by the MF dust model and its correlation with \ion{H}{i} (black solid line). We compare this to the opacities that would be inferred from the \ion{H}{i}-correlated dust SED derived by \citet{Planck_Int_XVII} assuming dust had a single temperature of 10, 15, 20, and 25\,K (red, green, blue, and black error bars, respectively). In addition to providing constraints on dust models, these opacities also highlight the sensitivity of determinations of the dust column to assumptions on the dust model (e.g., number of dust components, dust temperature).}
	\label{fig:mf_opacity}
\end{figure}

The MF reddening map is fundamentally a map of dust optical depth which is then converted to $E(B-V)$ assuming a linear relation. We can therefore employ this map to derive the dust optical depth per H atom implied by their two component dust model. Performing a linear fit analogous to that in Figure~\ref{fig:ebv_vs_nhi}, we find $\tau_{545}/N_\ion{H}{i} = \left(4.46\pm0.01\right)\times10^{-26}$\,cm$^2$. This corresponds to an opacity of

\begin{equation}
\kappa_{545} = 2.67 \frac{0.01}{M_{\rm d}/M_{\rm H}}\ {\rm cm}^2\,{\rm g}^{-1}
~~~,
\end{equation}
where $M_{\rm d}/M_{\rm H}$ is the dust to H mass ratio. Assuming the MF best-fit two component dust model, the frequency-dependence of $\kappa_\nu$ is given by

\begin{equation}
\kappa_\nu = \frac{f_1\frac{q_1}{q_2} + \left(1-f_1\right)\left(\frac{\nu}{\rm 3000\,GHz}\right)^{\beta_2-\beta_1}}{f_1\frac{q_1}{q_2} + \left(1-f_1\right)\left(\frac{\rm 545\,GHz}{\rm 3000\,GHz}\right)^{\beta_2-\beta_1}}\left(\frac{\nu}{\rm 545\,GHz}\right)^{\beta_1}\kappa_{545}
~~~,
\end{equation}
where $f_1 =  0.0485$, $q_1/q_2 = 8.22$, $\beta_1 = 1.63$, and $\beta_2 = 2.82$. We plot this opacity in Figure~\ref{fig:mf_opacity} assuming $M_{\rm d}/M_{\rm H} = 0.01$. For comparison, we also plot the opacities that would be inferred from the \ion{H}{i}-correlated dust SED derived by \citet{Planck_Int_XVII} assuming a single dust component with a temperature of 10, 15, 20, and 25\,K. The differences between the curves highlight the sensitivity of the inferred dust column to the assumed dust model, including the number of components and temperature of each.

The FIR dust opacity is particularly useful in constraining dust models since the wavelengths are much longer than the characteristic grain sizes and, consequently, the opacity is independent of the grain size distribution. However, we caution that the dust opacity presented here is contingent on the MF dust model, which we demonstrate has systematic temperature-dependent biases in its predictions of reddening (see Section~\ref{sec:compare}). Joint modeling of the \ion{H}{i} and FIR data may significantly improve the fidelity of the recovered opacities.

\section{An \ion{H}{i}-based reddening map}
\label{sec:reddening_map}

While we have established a tight relation between dust reddening $E(B-V)$ and total \ion{H}{i} column density in low density gas, some scatter remains. A portion of this scatter may be due to imperfections in the SFD reddening estimates arising from errors in the dust temperature correction, spatial variability of the FIR dust opacity, and contamination by extragalactic FIR emission and the zodiacal light. We explore this further in Section~\ref{sec:compare}. 

On the other hand, some scatter may be due to errors in the \ion{H}{i}-based reddening estimates due to non-\ion{H}{i} gas along the line of sight and variations in the dust to gas ratio in the ISM. In this Section, we discuss the potential of \ion{H}{i} velocity information to mitigate these uncertainties and produce more accurate predictions of $E(B-V)$ than the column density alone. We begin by fitting the relationship between \ion{H}{i} and $E(B-V)$ in a large number (16) of \ion{H}{i} velocity bins, and find that it is well-approximated by a constant value of $E(B-V)/N_\ion{H}{i}$ for $|v_{\rm LSR}| < 90$\,km\,s$^{-1}$ and zero otherwise. We then employ this ``tophat'' model to produce a map of interstellar reddening in the 39\% of the sky with $N_\ion{H}{i} < 4\times10^{20}$\,cm$^{-2}$.

\subsection{Velocity Dependence of $E(B-V)/N_{\ion{H}{i}}$}
\begin{figure*}[tp]
    \includegraphics[width=\columnwidth]{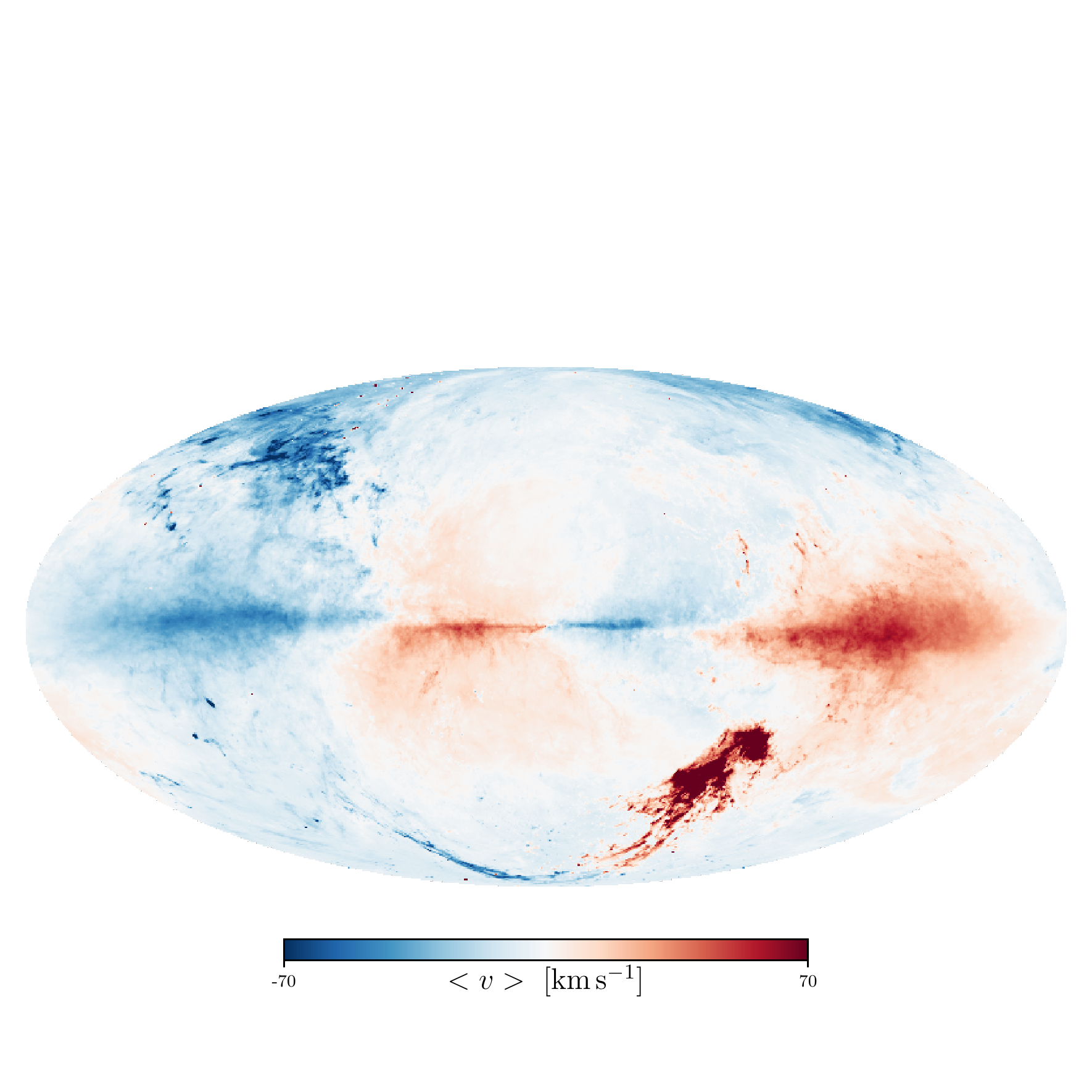}
	\includegraphics[width=\columnwidth]{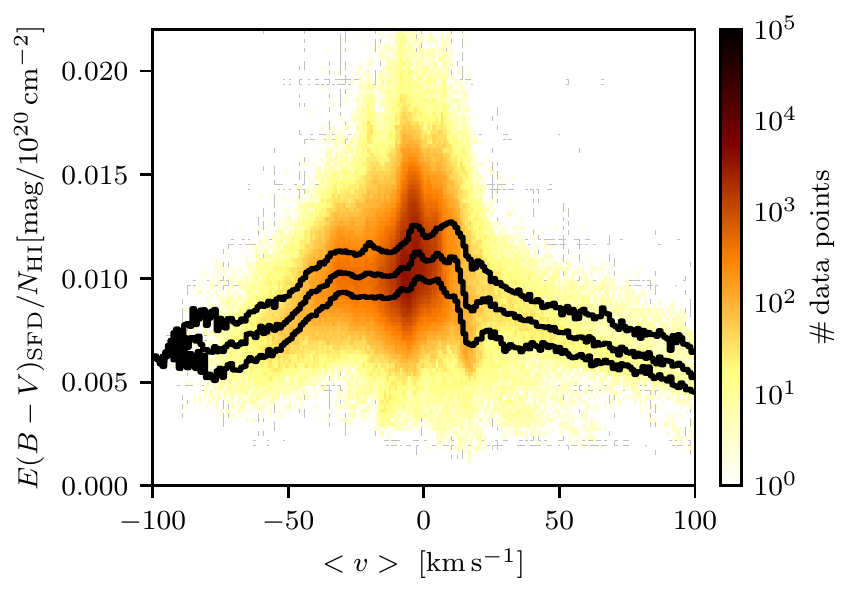}
	\caption{\textbf{Left:} Brightness temperature-weighted velocity map of the HI4PI data in Galactic coordinates, illustrating the large-scale features such as the Galaxy's rotation, the Magellanic system (high positive velocity feature in the southern sky) and the HVC complex C (high negative velocities in the top left). Without the binning scheme, all these very different structures would be treated equally in terms of their $N_{\ion{H}{i}}/E(B-V)$ ratio. \textbf{Right:} For each pixel with $N_\ion{H}{i} < 4\times10^{20}$\,cm$^{-2}$, we plot the measured reddening against the $T_{\rm B}$-weighted velocity $\langle v \rangle$ (see Equation~\ref{eq:nh_v}). The contour lines indicate 25, 50, and 75\% of the data points. The decline of $E(B-V)/N_\ion{H}{i}$ with increasing $|\langle v \rangle|$ is consistent with the presence of high velocity, low metallicity gas. We note that $<v>$ is a weighted average of $v_{\rm LSR}$ over all gas in each pixel and thus some of the high-velocity features are suppressed.}
	\label{fig:ebv_vs_v}
\end{figure*}

The linearity of the $N_\ion{H}{i}$-$E(B-V)$ relationship is fundamentally limited by spatial variations in the dust to gas ratio. These variations account for at least part of the 5\,mmag intrinsic scatter found between the SFD reddenings and the corresponding \ion{H}{i} column densities in Figure~\ref{fig:ebv_vs_nhi}. Some variations in the dust to gas ratio are traced by the $\ion{H}{i}$ velocity $v$. For instance, intermediate- and high-velocity clouds \citep{Wakker+vanWoerden_1997} are known to be deficient in dust relative to the low-velocity ISM \citep{Boulanger+etal_1996, Reach+etal_1998, Planck_2011_XXIV}, primarily due to their origin in the Galactic halo and beyond \citep{Wakker+vanWoerden_1997, Putman+etal_2012, Fraternali+etal_2015}. Further, there exist large-scale structures in the Galaxy with characteristic velocities and metallicities which can be identified in velocity space, such as the Magellanic Clouds and Magellanic Stream \citep{Nidever+etal_2008} as illustrated in the left panel of Figure~\ref{fig:ebv_vs_v}. Finally, many extragalactic \ion{H}{i} sources, which clearly do not contribute to Galactic reddening, can be identified by their high velocities and their shapes in position-velocity space.

We illustrate the systematic dependence of $E(B-V)/N_\ion{H}{i}$ with $v$ in Figure~\ref{fig:ebv_vs_v}. In the right panel we plot the observed $E(B-V)/N_\ion{H}{i}$ for all pixels with $N_\ion{H}{i} < 4\times10^{20}$\,cm$^{-2}$ against a measure of the gas velocity in that pixel. Since the \ion{H}{i} emission in each pixel generally arises from gas at various velocities, we compute an averaged velocity $\langle v \rangle$ weighted by the \ion{H}{i} brightness temperature at each $v$, i.e.,

\begin{equation}
\label{eq:nh_v}
\langle v \rangle \equiv \frac{\sum_v v T_{\rm B}(v)}{\sum_v T_{\rm B}(v)}
~~~,
\end{equation}
where $T_{\rm B}(v)$ is the brightness temperature of gas in that pixel having velocity $v$.

Figure~\ref{fig:ebv_vs_v} indicates two distinct, broad gas populations with characteristic dust to gas ratios, the first corresponding to low density low-velocity clouds and the second to medium/high density intermediate- and high velocity clouds. The pixels dominated by low velocity gas ($-30\,{\rm km}\,{\rm s}^{-1} \lesssim \langle v \rangle \lesssim 15\,{\rm km}\,{\rm s}^{-1}$) have an approximately constant value of $E(B-V)/N_\ion{H}{i}$ comparable with the $1.1\times10^{-22}$\,cm$^2$\,mag found in Section~\ref{sec:hi_ebv}. However, as high-velocity gas composes an increasingly large fraction of the total gas column (i.e. as $|\langle v \rangle |$ increases), the reddening per H atom decreases by roughly a factor of two, suggesting little reddening associated with the high-velocity gas. This is consistent with the known metal deficiency of intermediate- and high-velocity clouds.

In the following sections, we investigate the functional form of the dependence of $E(B-V)/N_\ion{H}{i}$ on $v$, first by dividing $N_\ion{H}{i}$ into a discrete velocity bins (Section~\ref{subsec:model}) and then by approximating the resulting relation with a simple tophat filter in velocity (Section~\ref{subsec:tophat}).

\subsection{Velocity-binned Model}
\label{subsec:model}
\begin{figure}[tp]
	\includegraphics[width=\columnwidth]{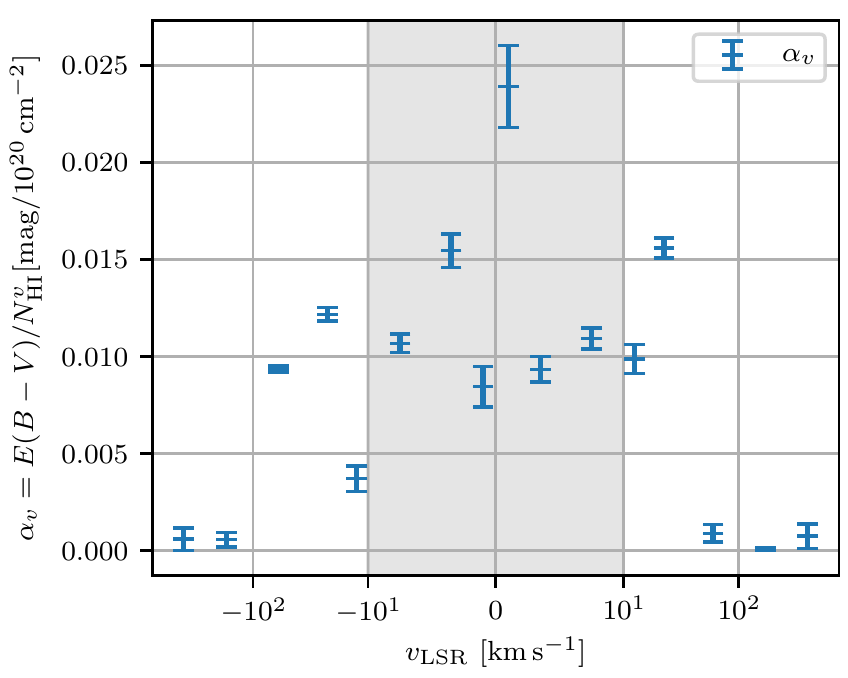}
	\caption{Fitted $\alpha_v \equiv E(B-V)/N^v_{\ion{H}{i}}$  (see Equation~\ref{eq:full_model}) as a function of radial velocity $v$. The error bars indicate the marginalized $1\sigma$ uncertainties, equivalent to the standard deviation for a Gaussian posterior. The velocity axis is scaled linearly over the shaded range and logarithmically at absolute velocities larger than $10$\,km\,s$^{-1}$. We note that the $\alpha_v$ tend to cluster around the value 1.1$\times10^{-22}$\,cm$^2$\,mag for $|v| \lesssim 100$\,km\,s$^{-1}$ and that the plotted errors are purely statistical. The scatter in the $\alpha_v$ and the varying size of the error bars are likely a reflection of inhomogeneities in the ISM due, for instance, to structures like low column density molecular clouds and the Magellanic Stream.}
	\label{fig:alpha_velocity_relation}
\end{figure}

To take advantage of the \ion{H}{i} velocity information, we model the SFD dust reddening $E(B-V)_i$ in a sky pixel $i$ as

\begin{equation}
	E(B-V)_i = \sum_v \alpha_{v} N^v_{\ion{H}{i},i} + \beta
	\label{eq:full_model}
~~~,
\end{equation}
where $N^v_{\ion{H}{i},i}$ is the \ion{H}{i} column density in pixel $i$ in velocity bin $v$, $\alpha_v$ is the reddening per H-atom for gas in velocity bin $v$, and $\beta$ is a constant offset. We employ a total of 16 velocity bins with finer binning at low absolute velocities to account for the complexity of the structures in the Galactic disk, as illustrated in Figure~\ref{fig:alpha_velocity_relation}. We seek the best fit values of $\beta$ and the 16 $\alpha_v$. We find that our conclusions are not sensitive to the exact choice of bin numbers or binning scheme (e.g. linear, log-spaced, or non-uniform bins).

For the vector of fit parameters $\theta = \lbrace\alpha_{v}, \beta, \sigma_{\rm tot}\rbrace$ and the data vector $\mathcal{D} = \lbrace N_{\ion{H}{i}}^v, E(B-V)_{\rm SFD}\rbrace$, the full posterior probability density function is:
\begin{equation}
	p(\theta | \mathcal{D}) \propto \mathcal{L}(\mathcal{D} | \theta) p(\theta), \\
	\label{eq:posterior}
\end{equation}
and the likelihood is given by
\begin{equation}
	\mathcal{L}(\mathcal{D} | \theta) \sim \mathcal{N}\left(E(B-V)_{\rm SFD} - E(B-V)_{\rm model} ,\sigma_{\rm tot}^2\right).
    \label{eq:likelihood}
\end{equation}
where $\sigma_{\rm tot}$ is the total unknown scatter of the relation, encompassing both the intrinsic scatter in the relationship and the observational uncertainties in the $E(B-V)$ and \ion{H}{i} maps, i.e., $\sigma^2_{\rm tot} = \sigma^2_{\rm intr} + \sigma^2_{\rm data}$. The measurement errors on $N_\ion{H}{i}$ are small ($\lesssim 5\%$) and we ignore them here. The SFD maps are not provided with associated uncertainties, and so we cannot separate their contribution to $\sigma^2_{\rm tot}$ from that due to the intrinsic scatter.

We adopt the following priors on the model parameters:
\begin{align}
    \label{eq:prior}
    \alpha_v & = \tan\phi_v\nonumber\\
    p(\phi_v) & = \mathrm{Uniform}\left(\phi_v, 0, \pi/2\right)\\
    p(\ln\sigma_{\rm tot}) & = \mathrm{Uniform}\left(\ln\sigma_{\rm tot}, -7, 1\right)\nonumber \\
    p(\beta) & = \mathcal{N}\left(\beta, 0, 0.1\right) \nonumber
\end{align}
For the reddening per neutral hydrogen atom $\alpha_v$, we sample uniformly in $\phi_v = \arctan(\alpha_v)$ to avoid a bias towards larger values and we constrain $\alpha_v$ to positive values. For the scatter $\sigma_{\rm tot}$, we employ a broad scale-invariant prior. The offset $\beta$ is centered on zero with a standard deviation of 0.1\,mag, following the results of previous studies \citep{Schlafly+etal_2014, Meisner+Finkbeiner_2015}. Sampling is performed with the \texttt{emcee}\footnote{\url{http://dan.iel.fm/emcee}} package \citep{Foreman-Mackey+etal_2013}.

In agreement with the simple linear relation fit in Figure~\ref{fig:ebv_vs_nhi}, we find a very small offset $\beta = \left(0.25\pm 0.12\right)\, \rm mmag$, indicating excellent zero point agreement between the SFD reddening and the \ion{H}{i} data. The best fit $\sigma_{\rm tot} =\left(5.1\pm 0.1\right)\, \rm mmag$ also agrees with the $5.7$\,mmag intrinsic scatter inferred from the simple linear fit. 

The derived values of the reddening per H-atom for each velocity bin are shown in Figure~\ref{fig:alpha_velocity_relation}. As expected from the trend observed in Figure~\ref{fig:ebv_vs_v} (right), the ratio drops for absolute radial velocities larger than approximately 100$\,\rm km\,s^{-1}$. With the exception of a few outliers, the $\alpha_v$ in the lower velocity bins are clustered around $0.011\,\rm mag/10^{20}\,cm^{-2}$, the $E(B-V)$/$N_\ion{H}{i}$ value obtained from the linear fit.

The high value for the bin in between 0 and 2$\,\rm km\,s^{-1}$ can be explained by a high latitude filament that, despite its low \ion{H}{i} column density, contains molecular hydrogen. It is associated with a larger molecular intermediate-velocity cloud \citep{Magnani+Smith_2010, Roehser+etal_2016} and accounts for a large fraction of the pixels having most of their gas in that velocity bin.

Inhomogeneities in the dust to gas ratio within the Galactic ISM, such as the presence of H$_2$ at low column densities or large scale structures like the Magellanic Stream, therefore account for much of the variation in the fit $\alpha_v$ values and the non-uniformity in the error bars. Indeed, deviations from the expected $E(B-V)/N_\ion{H}{i}$ can be used to {\it locate} such structures, as we demonstrate above. However, it would not be appropriate to extrapolate the observed reddening per H atom in these structures to diffuse gas at the same velocity elsewhere in the Galaxy. We therefore seek a model which is complex enough to account for the systematic decline of the reddening per H atom with increasing velocity while simple enough such that anomalous variations are averaged out.

\subsection{Tophat Model}
\label{subsec:tophat}
\begin{figure}[tp]
	\includegraphics[width=\columnwidth]{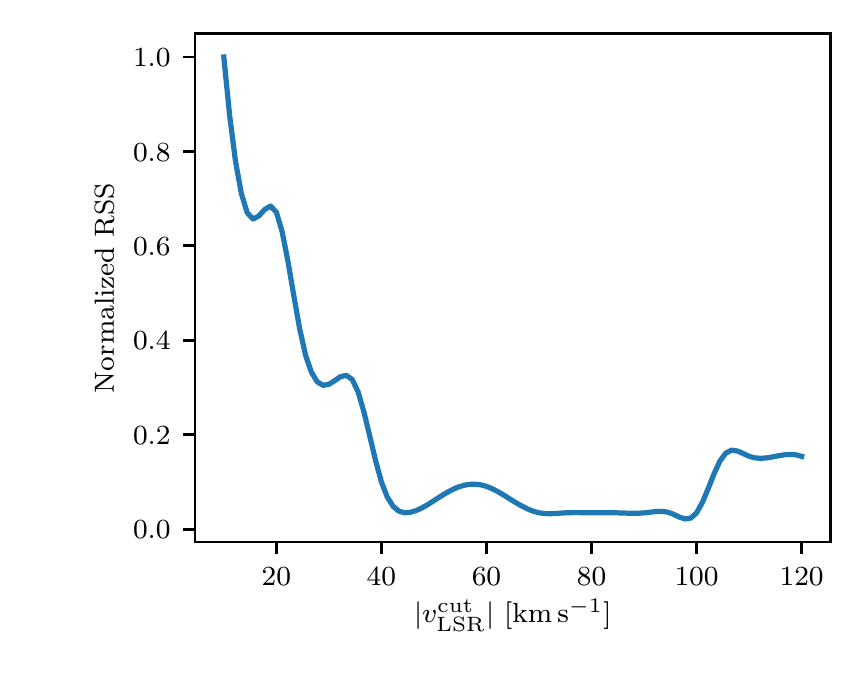}
    \caption{Residual sum of squares (RSS), normalized to the maximum, for different velocity ranges of our tophat model (Section~\ref{subsec:tophat}) such that only gas with $|v| < v_{\rm LSR}^{\rm cut}$ is assumed to correlate with reddening. The minimum between 70 and 100\,$\rm km\,s^{-1}$ corresponds to the transition from local gas to dust-poor HVCs for which the $E(B-V)/N_{\ion{H}{i}}$ ratio is much smaller (see also Figure~\ref{fig:differences_simple_tophat}).}
	\label{fig:tophat_search}
\end{figure}

\begin{figure*}[tp]
	\includegraphics[width=\textwidth]{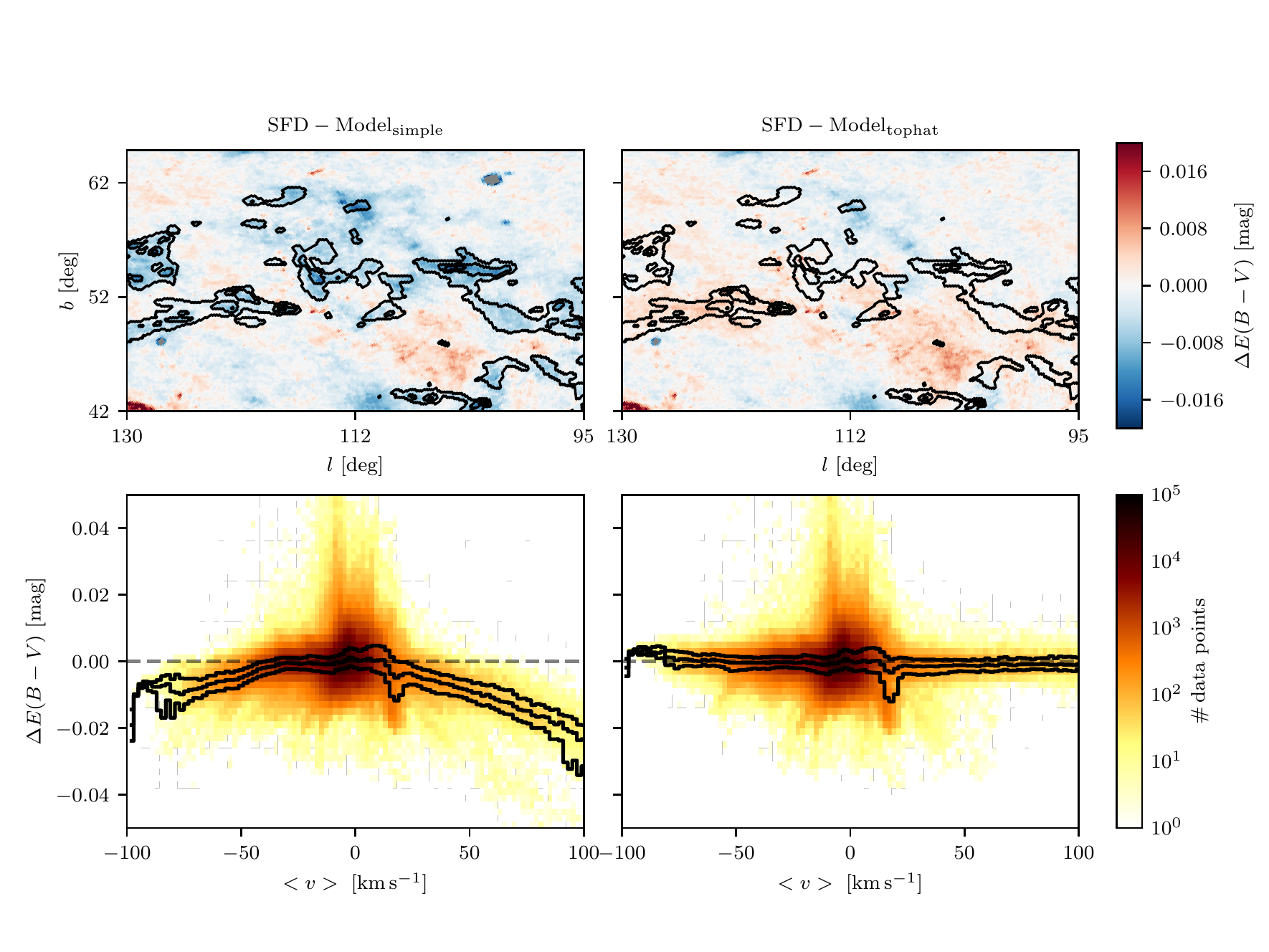}
    \caption{\textbf{Top:} Residuals between the SFD map and our models in the direction of high-velocity cloud complex C. Contours indicate the \ion{H}{i} column density for velocities between $-400\,\rm km\,s^{-1}$ and $-90\,\rm km\,s^{-1}$ beginning at $4\times 10^{19}\,\rm cm^{-2}$ and increasing in steps of $4\times 10^{19}\,\rm cm^{-2}$. Left are the residuals for the simple model without any velocity limit, right the residuals for the tophat model. \textbf{Bottom:} Residual reddening (SFD - Model) as a function of $<v>$ (Eq.~\ref{eq:nh_v}) for the simple model (left) and the tophat model (right). The contour lines indicate 25, 50, and 75\% of the data points. The tophat model effectively removes most systematic trends with velocity in the residuals.}
    \label{fig:differences_simple_tophat}
\end{figure*}

In this Section, we develop a model intermediate in complexity to a simple linear relationship between $N_\ion{H}{i}$ and $E(B-V)$ and fine binning in radial velocity (Eq.~\ref{eq:full_model}). We consider a simple ``tophat'' model with a constant $E(B-V)/N_\ion{H}{i}$ for gas with $|v_{\rm LSR}| < v_{\rm LSR}^{\rm cut}$ and no reddening associated with gas outside that velocity range. This model is motivated by the fact that the total \ion{H}{i} column density scales very tightly with reddening for low densities (Sect.~\ref{sec:hi_ebv}). Furthermore, we note that the $E(B-V)/N_{\ion{H}{i}}$ ratio for the different velocity bins in Figure~\ref{fig:alpha_velocity_relation} scatters around the mean value of $0.011\,\rm mag/10^{20}\,cm^{-2}$ for low and intermediate velocities and is very close to zero for high-velocity gas. 

To select the value of $v_{\rm LSR}^{\rm cut}$ quantitatively, we fit a simple linear relation between $N_\ion{H}{i}$ and $E(B-V)$, where the $N_\ion{H}{i}$ in each pixel is restricted to gas having velocity $|v_{\rm LSR}| < v_{\rm LSR}^{\rm cut}$. In Figure \ref{fig:tophat_search}, we plot the normalized residual sum of squares (RSS) of the fit as a function of $v_{\rm LSR}^{\rm cut}$. We find a broad minimum from $\sim$70--100\,$\rm km\,s^{-1}$, consistent with typical HVC velocities \citep{Wakker+vanWoerden_1997}. We therefore adopt $v_{\rm LSR}^{\rm cut} = 90\,\rm km\,s^{-1}$.

To illustrate the advantage of this limit in radial velocity, we show in Figure~\ref{fig:differences_simple_tophat} (top) the residual reddening for both models with respect to SFD in the direction of high-velocity cloud complex C \citep[e.g.][]{Fox+etal_2003, Wakker+etal_2007, Fraternali+etal_2015}. The contours indicate the \ion{H}{i} column density for velocities lower than $-90\,\rm km\,s^{-1}$. The simple model without any constraints in radial velocity falsely attributes reddening to the low metallicity, low-reddening, high-velocity gas and therefore overestimates the true reddening along these lines of sight. The simple tophat filter in radial velocity largely overcomes this problem.

\begin{figure}[tp]
	\includegraphics[width=\columnwidth]{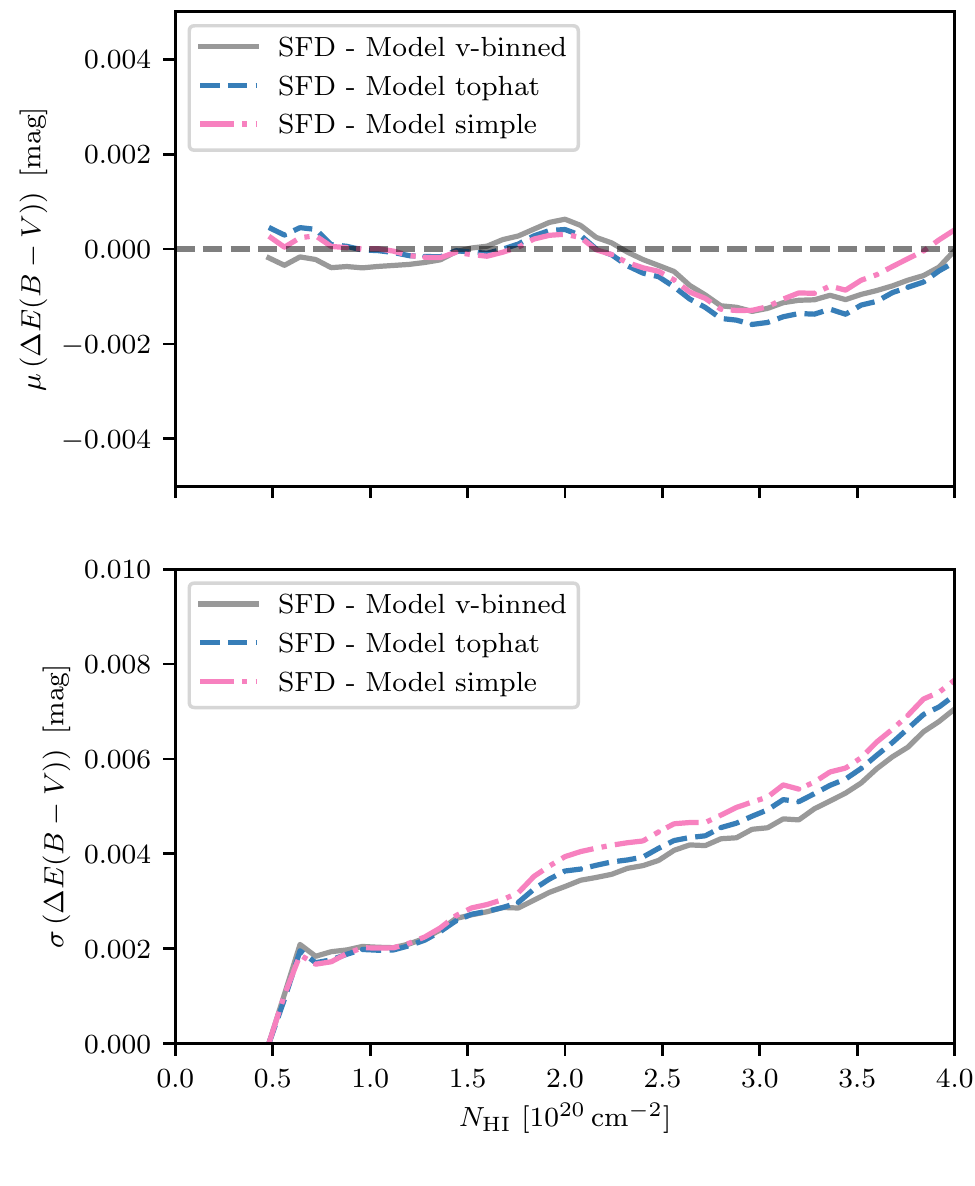}
	\caption{Median difference ({\bf top}) and standard deviation ({\bf bottom}) of the residuals with respect to the SFD map as a function of \ion{H}{i} column density for our various models. ``v-binned'' refers to the model with sixteen velocity bins described in Section~\ref{subsec:model}, tophat refers to a single \ion{H}{i} column density map with $|v_{\rm LSR}| < 90\,\rm km\,s^{-1}$, and simple refers to a linear relationship between $E(B-V)$ and $N_\ion{H}{i}$ with no constraints for the velocity.}
	\label{fig:residual_stddev_intern}
\end{figure}

We further assess the tophat model in Figure~\ref{fig:differences_simple_tophat} (bottom), where we compare the residuals of the simple and tophat models as a function of $<v>$. While the simple model markedly overestimates the redding at large $|<v>|$, the $<v>$ dependence is nearly completely removed in the tophat model. We do note, however, that the complexity of the ISM at local velocities leads to outliers at low $|<v>|$, though only a small fraction of the data is affected.

A quantitative comparison of the different model flavors is presented in Figure~\ref{fig:residual_stddev_intern}, where we compare the median agreement and the scatter of each of these models with respect to SFD as a function of $N_\ion{H}{i}$. For all models, the median agreement with the SFD data is very good with deviations of less than 2\,mmag at fixed $N_\ion{H}{i}$, and with minimal (sub-mmag) differences between the models. For the standard deviation of the residuals, we find that the full model with a large number of velocity bins performs best, but the differences to the tophat and the very simple model are only of the order of 1\,mmag. The simple linear fit model, while having a slightly smaller mean bias than the other models, has the largest the scatter at fixed $N_\ion{H}{i}$, indicating that the relevant physics is not being entirely captured.

\begin{figure*}[tp]
	\includegraphics[width=\textwidth]{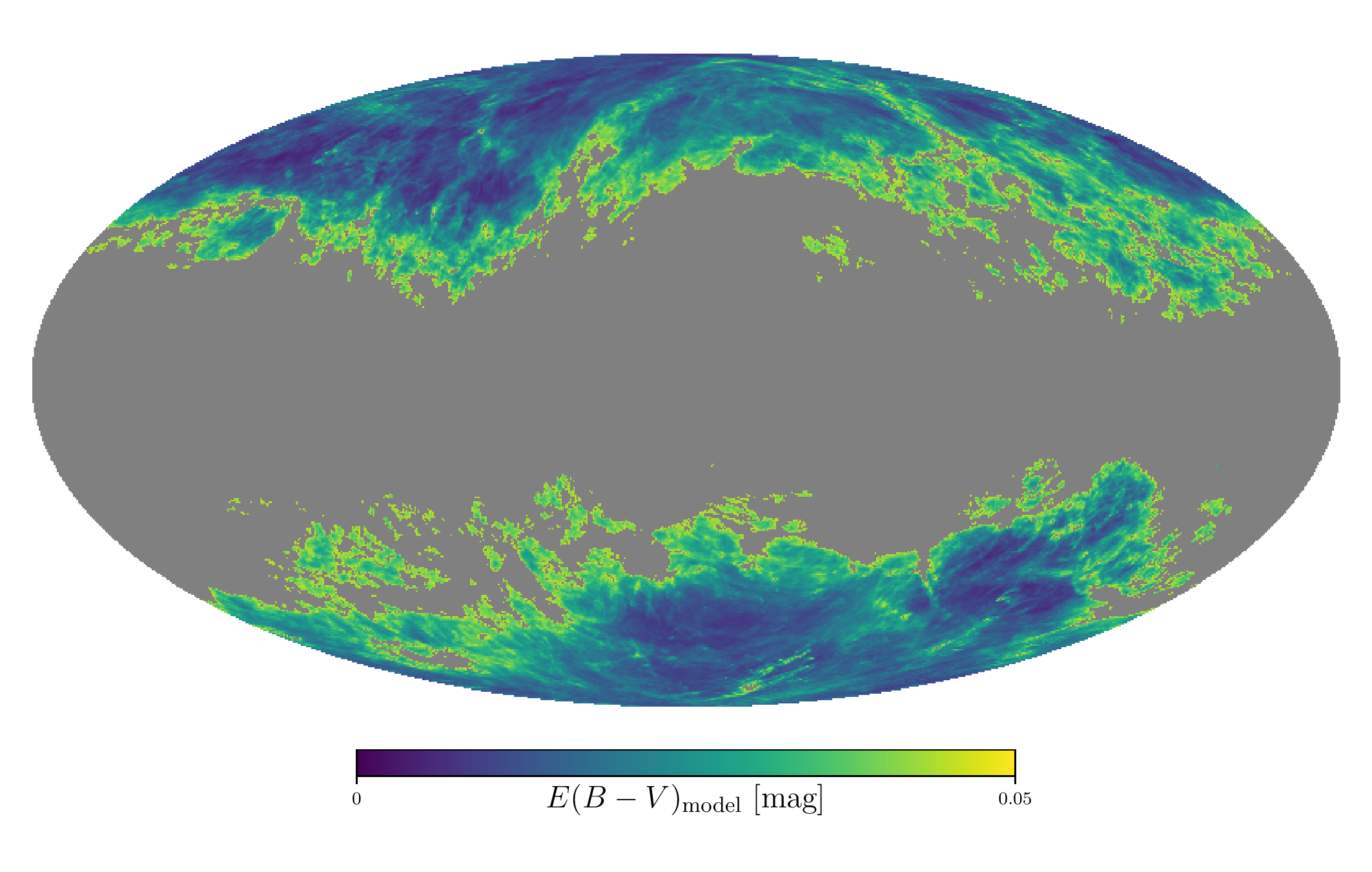}
	\caption{The $E(B-V)$ map derived in this work by correlating \ion{H}{i} emission with the SFD reddening map. We find that \ion{H}{i} gas with velocity $|v_{\rm LSR}| > 90$\,km\,s$^{-1}$ is uncorrelated with Galactic reddening. See Section~\ref{sec:reddening_map} for details on the model.}
	\label{fig:model}
\end{figure*}

In summary, we find that for the low density-sky, a simple linear scaling with total \ion{H}{i} column density already provides a rather good fit to the SFD reddenings. By limiting the velocity range to exclude the high-velocity gas, we can improve this model in a straightforward and physically-motivated manner. The addition of the complex binning scheme in radial velocities helps to account for some of the $E(B-V)/N_{\ion{H}{i}}$-variations at low velocities, but also renders the model more susceptible to fitting biases in the SFD map rather than true variations in $E(B-V)/N_{\ion{H}{i}}$. Additionally, the improved agreement with the SFD map relative to the tophat model is marginal, as illustrated in Figure~\ref{fig:residual_stddev_intern}. Thus, we construct our final reddening map based on the tophat model as it offers a good balance between accuracy and simplicity. The derived model parameters are combined with the \ion{H}{i} data an angular resolution of $16.1'$ and sampled on a grid with $\rm N_{side}=1024$. The resulting reddening map is illustrated in Figure~\ref{fig:model}.

\subsection{\ion{H}{i} Systematics}
\begin{figure*}[tp]
	\includegraphics[width=\textwidth]{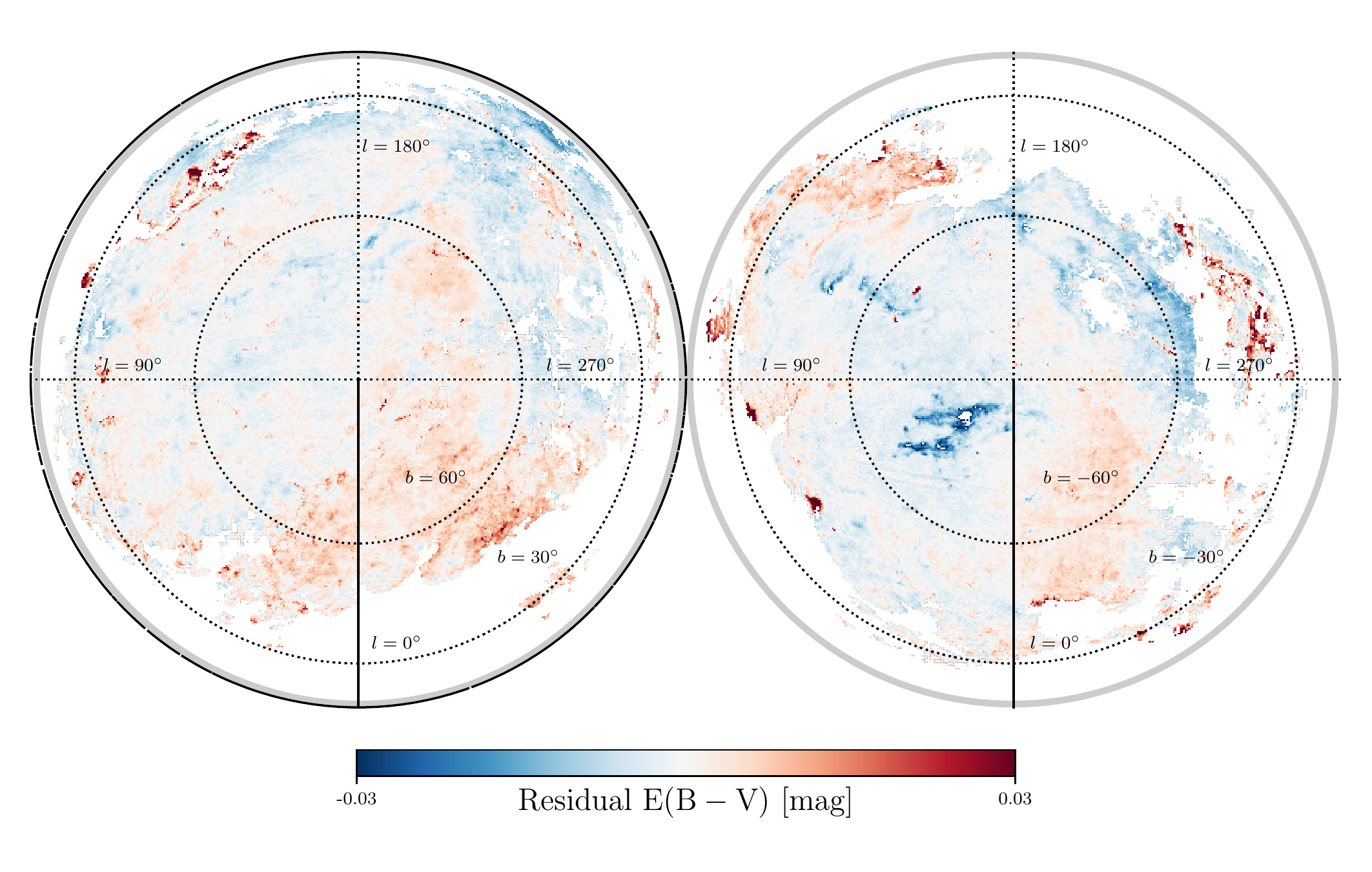}
	\caption{Large-scale $E(B-V)$ residuals (SFD - model) in zenith equal area projection and in Galactic coordinates. The northern Galactic hemisphere is shown left, the southern hemisphere right. Galactic longitude $l=0^{\circ}$ connects the center and the bottom of each hemisphere and $l$ increases clockwise for the northern hemisphere and counter-clockwise for the southern hemisphere.}
	\label{fig:mollweide_residual}
\end{figure*}

The simple tophat model relating $N_\ion{H}{i}$ to $E(B-V)$ is excellent at reproducing the observed reddenings on the mean. In this section, we assess whether there exist other systematic residuals with the \ion{H}{i} data.

The large-scale residuals $E(B-V)_{\rm SFD} - E(B-V)_{\rm model}$ are shown in Figure~\ref{fig:mollweide_residual}. Although the overall amplitude of the residuals is small ($\sigma \simeq 5\,$mmag), some large scale structures are seen, particularly toward the Galactic plane and Galactic Center. This may be due to H$_2$ formation in gas close to our column density limit of $4\times 10^{20}\,\rm cm^{-2}$ such that the \ion{H}{i} emission no longer traces the full gas column (see also Figure~\ref{fig:ebv_params}. However, some of the large scale features may be intrinsic to the SFD map, as we argue in Section~\ref{sec:compare}.

\begin{figure*}[tp]
	\includegraphics[width=\columnwidth, viewport= 0 0 200 180, clip]{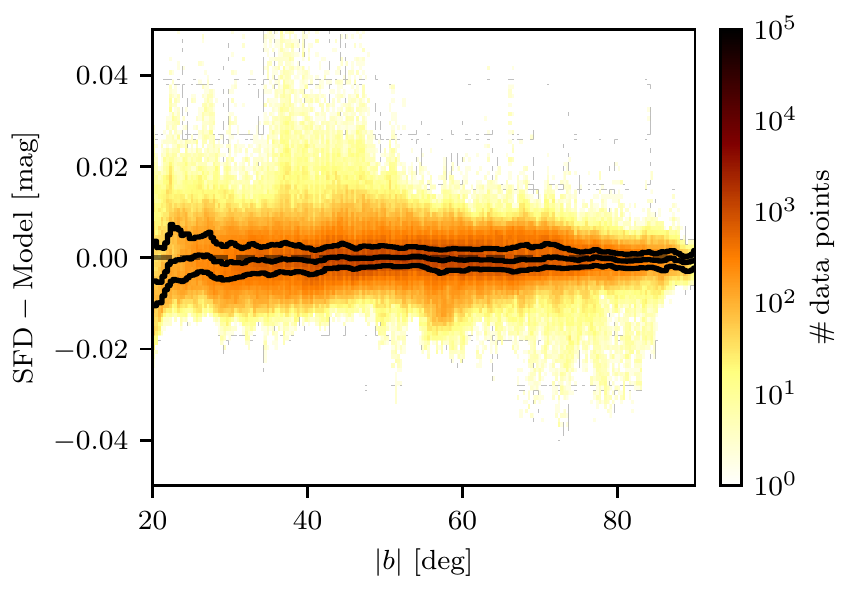}
    	\includegraphics[width=\columnwidth, viewport= 40 0 240 210, clip]{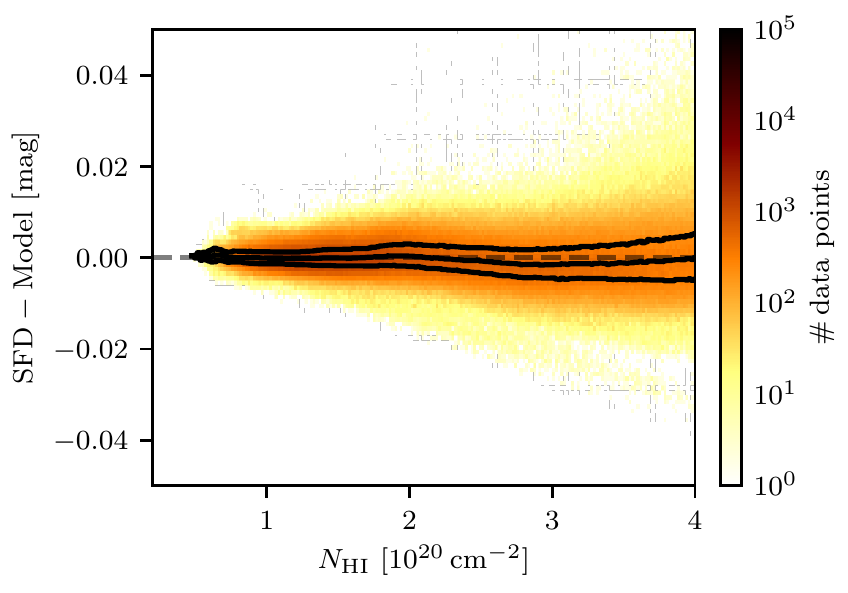}
	\caption{Residual reddening (SFD - Model) as a function of absolute Galactic latitude (\textbf{left}) and of  \ion{H}{i} column density (\textbf{right}). The contour lines indicate 25, 50, and 75\% of the data points.}
	\label{fig:residuals_vs_lat_and_nhi}
\end{figure*}

It is plausible that there is a gradient in the dust to gas ratio with Galactic latitude, which would induce a bias in our predicted reddenings. Further, the Galactic ISM becomes increasingly complex closer to the disk, so it is also plausible that H$_2$ contamination is greater at low $|b|$. However, as illustrated in Figure~\ref{fig:residuals_vs_lat_and_nhi} (left), we find little variation in the model residuals with $|b|$. While the residuals do tighten slightly towards the Galactic poles and show some structure below $|b| \simeq 25^\circ$ where the number of data points becomes small, both effects are minor. Any systematic changes in the reddening per H atom with Galactic latitude must therefore be minor, at least for low column density gas.

In Figure~\ref{fig:residuals_vs_lat_and_nhi} (right), we plot the residuals (SFD - Model) against the \ion{H}{i} column density, which illustrates similar trends as seen in Figure~\ref{fig:residuals_vs_lat_and_nhi} (left). For data below the threshold of $N_{\ion{H}{i}} < 4\times 10^{20}\,\rm cm^{-2}$ (corresponding to $E(B-V) < 0.045\,\rm mag$), we find no systematic bias in the residuals. As expected, the relation between reddening and total \ion{H}{i} column density is very tight for the purely atomic phase of the dust-to-gas relation ($N_{\ion{H}{i}}\lesssim 2\times 10^{20}\,\rm cm^{-2}$). Above that value, the complexity of the relation increases and leads to a larger scatter. Close to our threshold of $N^{\rm threshold}_{\ion{H}{i}} = 4\times 10^{20}\,\rm cm^{-2}$, we observe a very minor skew in the residuals towards higher reddening values in the SFD data likely due to the presence of molecular gas.

Overall, the \ion{H}{i}-based reddenings reproduce the SFD map with high fidelity (scatter $\simeq 5$\,mmag) and with little to no discernible bias with $N_\ion{H}{i}$, $b$, or \ion{H}{i} velocity. We now investigate whether the observed residual structure is due to systematics in the dust-based maps, providing for the first time a stringent test of these maps at high resolution in the low column density limit.

\section{Comparison to Other Reddening Maps}
\label{sec:compare}

\begin{figure*}[tp]
	\includegraphics[width=0.33\textwidth, viewport=120 10 460 368, clip]{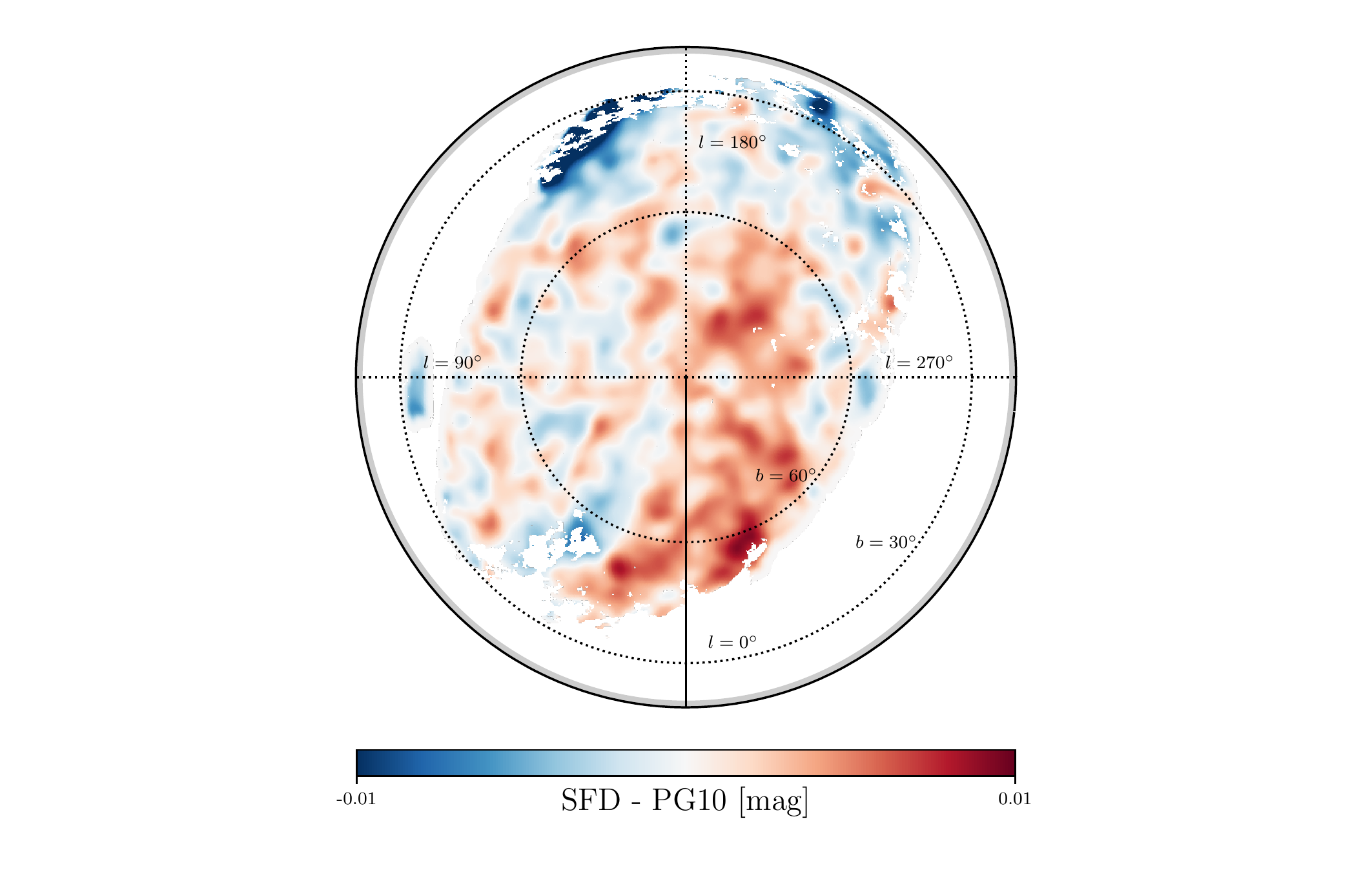}
    \includegraphics[width=0.33\textwidth, viewport=120 10 460 368, clip]{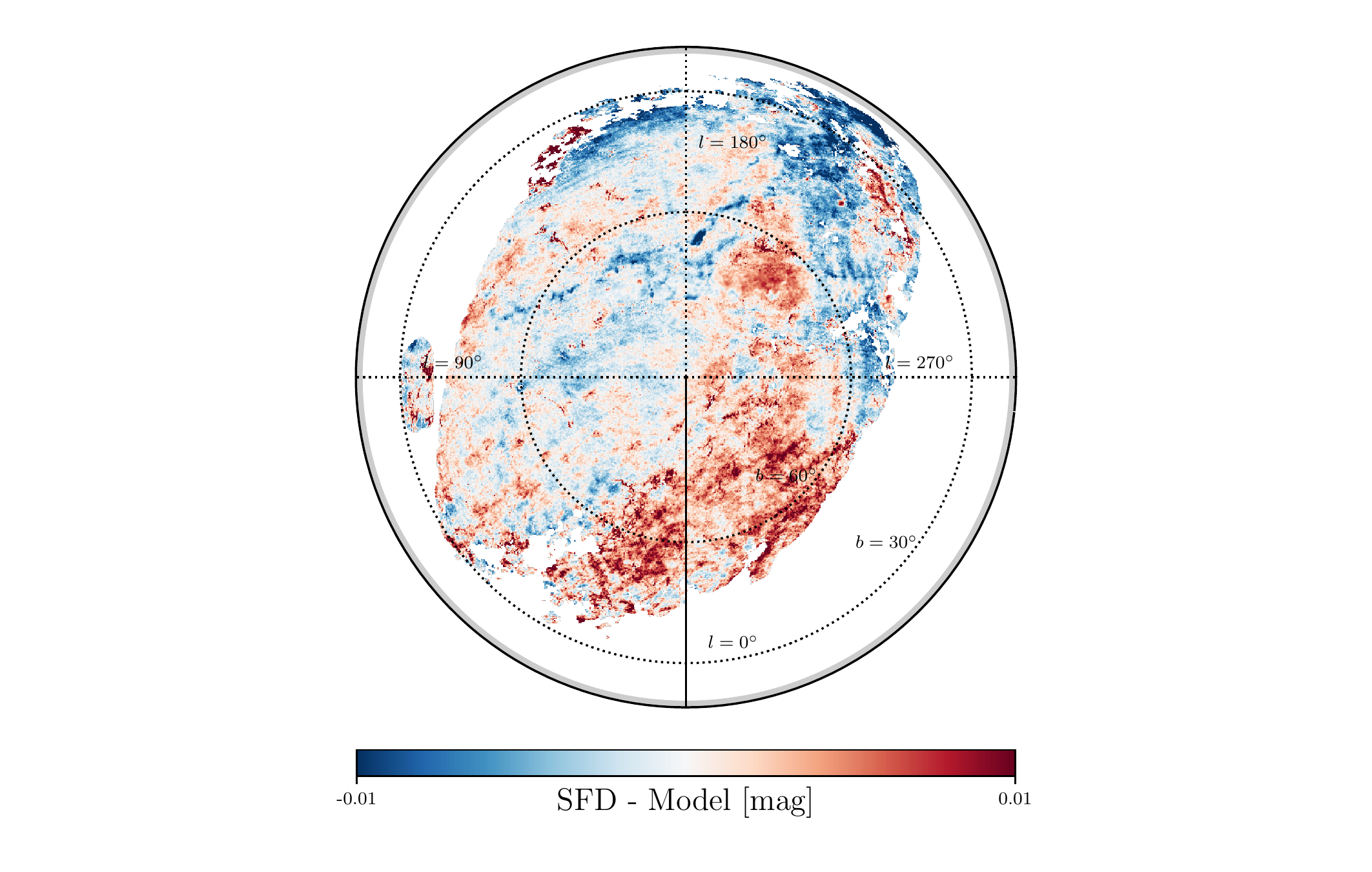}
    \includegraphics[width=0.33\textwidth, viewport=120 10 460 368, clip]{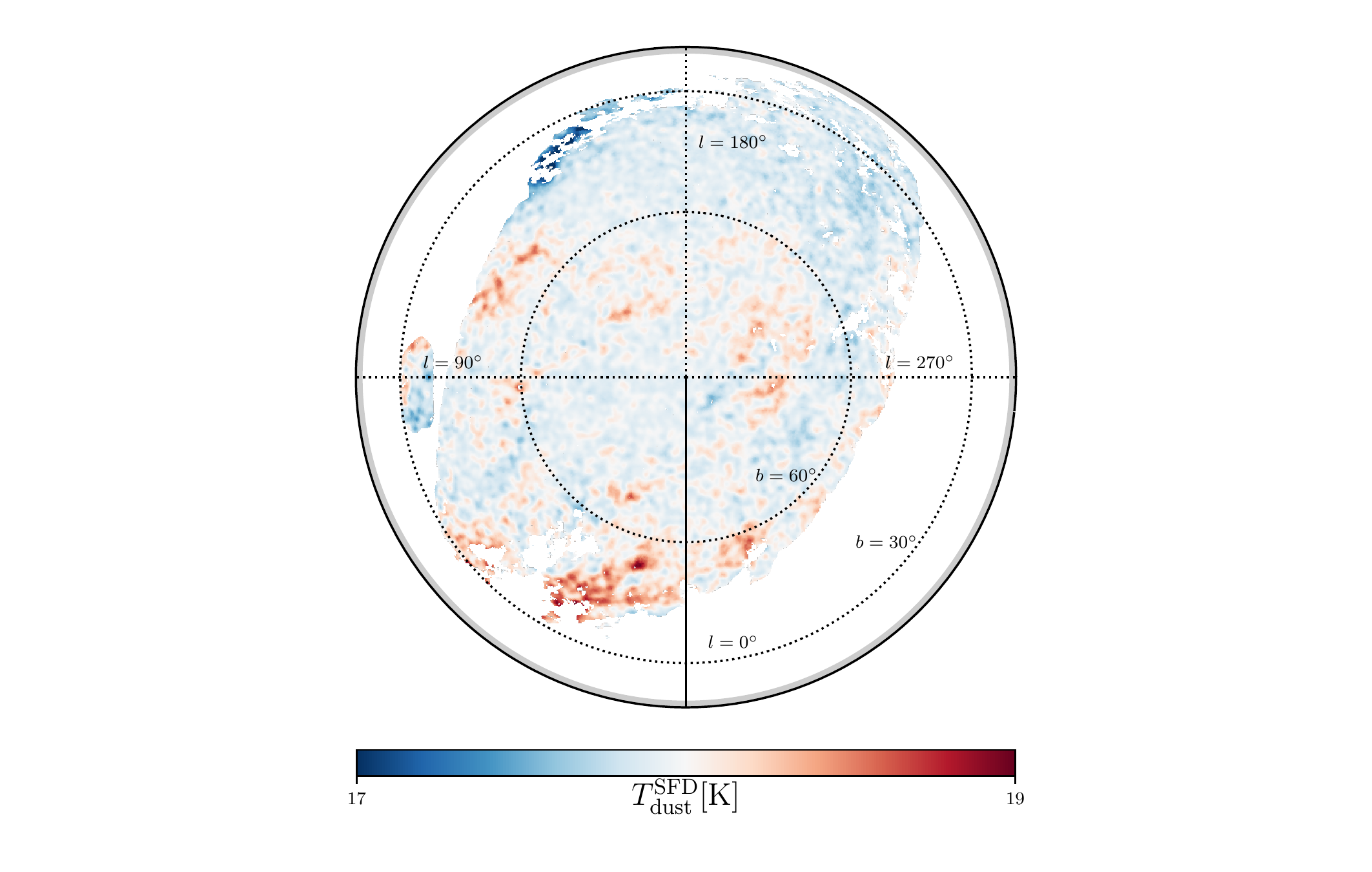}
	\caption{Correction to the SFD map by \citet{Peek+Graves_2010} based on optical observations of passive galaxies (\textbf{left}) and the differences between the SFD map and the \ion{H}{i}-based map derived in this work (\textbf{center}). Given the completely independent methods, the agreement between the two maps is striking. The \textbf{right} image shows the SFD dust temperature, which shows little correlation with the large scale structures in the other images. For all images, we show the zenith equal area projection of the northern Galactic hemisphere. Galactic longitude $l$ increases clockwise, $l=0^{\circ}$ points down.}
	\label{fig:pg10_comparison}
\end{figure*}

\begin{figure}[tp]
	\includegraphics[width=\columnwidth]{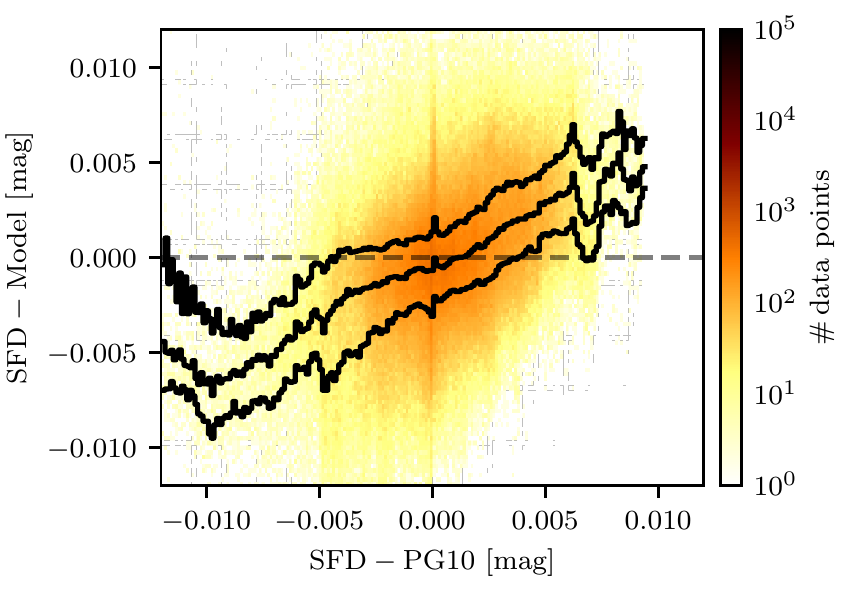}
	\caption{The differences between the SFD data and this model as a function of the correction presented by \citet{Peek+Graves_2010}. We find a clear correlation between the two even though these corrections to SFD were derived in orthogonal ways. The contour lines indicate 25, 50, and 75\% of the data points.}
	\label{fig:residuals_vs_pg10}
\end{figure}

In Section~\ref{sec:reddening_map}, we constructed an \ion{H}{i}-based reddening map in excellent overall agreement with SFD, with median bias less than 1\,mmag and a dispersion of $\simeq 5$\,mmag. However, Figure~\ref{fig:mollweide_residual} suggests some degree of large scale structure in the residuals. While this can be attributed in part to Galactic structures with anomalous reddening per H atom, such as in regions with significant H$_2$ even at low column density, these structures may also be due to systematics in the SFD map itself. Indeed, \citet{Peek+Graves_2010} found large scale residuals when comparing the SFD reddenings to direct reddening measurements toward passive galaxies. In this Section, we employ our \ion{H}{i}-based reddening map as a stringent test of existing reddening maps in the low column density limit and demonstrate that much of the disagreement with the \ion{H}{i}-based reddenings can be explained by systematics in the dust emission-based maps.

We begin by comparing our residuals with respect to the SFD map with the corrections to SFD derived by \citet{Peek+Graves_2010} based on observations of passive galaxies in Figure~\ref{fig:pg10_comparison}. Despite the orthogonal approaches in determining reddening and the much lower resolution of the \citet{Peek+Graves_2010} map (4$\overset{\circ}{.}$5 vs. $16\overset{'}{.}1$), we find striking qualitative agreement between the large scale structures in the residual maps.

A more quantitative comparison between our reddening map and that of \citet{Peek+Graves_2010} is presented in Figure~\ref{fig:residuals_vs_pg10}. To facilitate direct comparison, we first convolve our reddening map to a matching resolution of 4$^\circ$.5 and then compare the residuals of each map with respect to SFD. We find a positive correlation between the residuals at high significance, indicating that the \ion{H}{i} map is detecting the same discrepancies with SFD as the \citet{Peek+Graves_2010} map. We find that the \ion{H}{i} and \citet{Peek+Graves_2010} maps differ by only 4.4\,mmag on average, less than the 5.7\,mmag of intrinsic scatter found between the \ion{H}{i} and SFD maps in Figure~\ref{fig:ebv_vs_nhi}. Given the superior agreement with an external calibrator, we conclude that the \ion{H}{i}-based map is the more reliable predictor of reddening in the low column density regime.

\begin{figure*}[tp]
	\includegraphics[width=2\columnwidth]{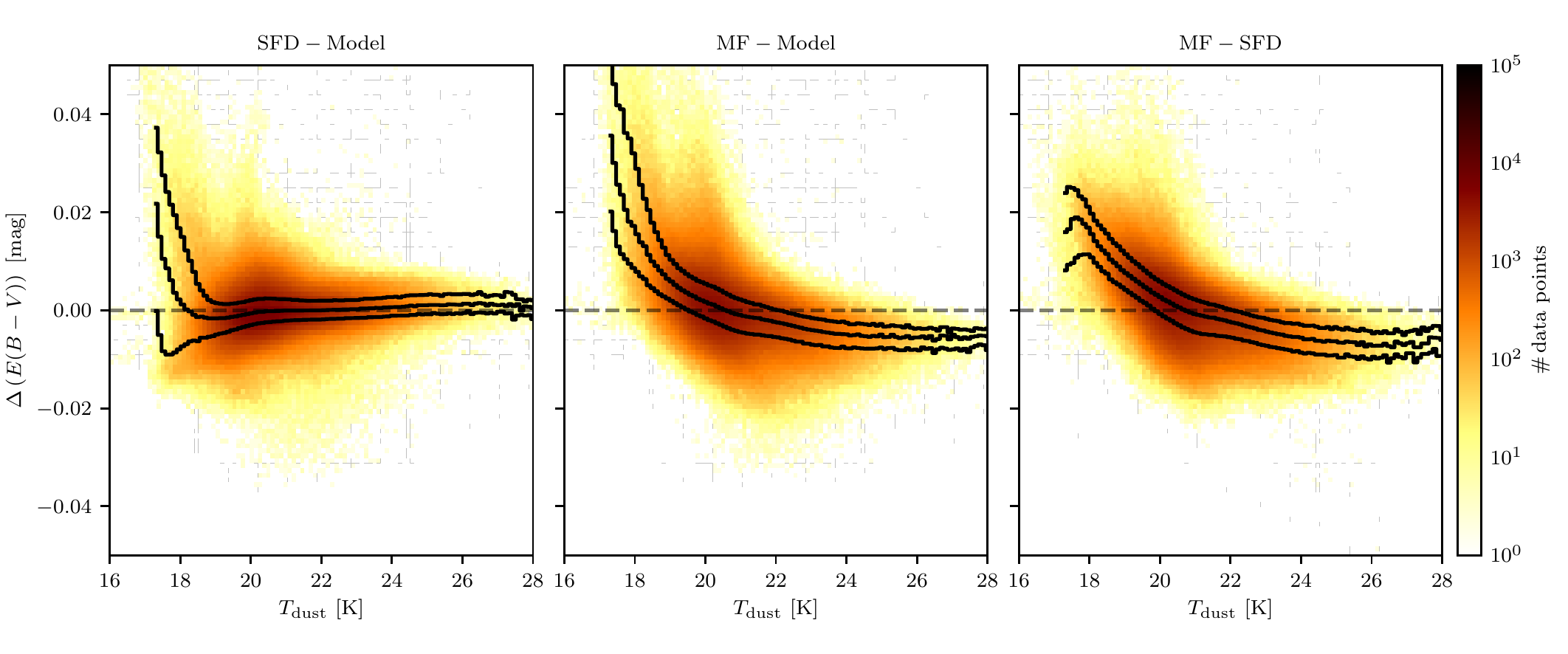}
	\caption{Reddening residuals with dust emission-based maps as a function of dust temperature. The contour lines indicate 25, 50, and 75\% of the data points. Systematic trends with dust temperature are evident in all panels.}
	\label{fig:residuals_vs_dust_temperature}
\end{figure*}

We investigate the influence of dust temperature in Figure~\ref{fig:residuals_vs_dust_temperature}, where we plot the residuals of different reddening maps with the \ion{H}{i}-based map as a function of dust temperature. We employ the dust temperature map derived by the \textit{Planck} foreground modeling pipeline \citep{Planck_2013_XI}, smooth it to an angular resolution of $16\overset{'}{.}1$, and down-sample it to a HEALPix $\rm N_{side} = 512$. The SFD residuals are relatively flat with structure emerging only at the low-dust temperature end. The very lowest dust temperatures are likely associated with molecular gas, so it is therefore difficult to ascertain whether this feature is due to bias in the SFD or \ion{H}{i} reddenings. We do note that residuals increase monotonically with dust temperature, but the overall effect is very slight. 

In contrast, there is a strong trend across the entire range of dust temperatures for our residuals with respect to the MF map. These residuals may originate in incorrect temperature corrections, CIB fluctuations, or both. In particular, the ratio of the IRAS 100\,$\mu$m intensity to the {\it Planck} 857\,GHz intensity was critical for determining the dust temperature in the \textit{Planck} analysis and is also sensitive to CIB fluctuations. As MF employed both {\it Planck} 857\,GHz and IRAS 100\,$\mu$m in their models, the $T_d$ we plot in Figure~\ref{fig:residuals_vs_dust_temperature} may be more sensitive to errors in MF than SFD, who relied on much lower resolution DIRBE data to make temperature corrections.

We explore this possibility in Figure~\ref{fig:residuals_vs_xmap} by plotting our residuals with SFD against the SFD temperature correction factor $X_{\rm SFD}$. This factor is defined such that the true DIRBE 100\,$\mu$m flux is $X_{\rm SFD}$ times what the flux would be if the dust were at a temperature of 18.2\,K. In practice, it is determined by the ratio of the DIRBE 100 and 240\,$\mu$m fluxes \citep[see][Section 2.4.2 for details]{Schlegel+Finkbeiner+Davis_1998}. Thus, large values of $X_{\rm SFD}$ generally correspond to low dust temperatures. As in Figure~\ref{fig:residuals_vs_dust_temperature}, we note discrepancies at very low dust temperatures (large $X_{\rm SFD}$) but otherwise little systematic variation. Additionally, we note little qualitative relationship between maps of $X_{\rm SFD}$ or the SFD dust temperature with our map of residuals (Figure~\ref{fig:pg10_comparison}). Thus, it does not appear that the dust temperature corrections are driving the large scale discrepancies between SFD and the \ion{H}{i} reddening map noted in Figure~\ref{fig:pg10_comparison} though it is true, as noted by \citet{Peek+Graves_2010}, that the region with the greatest reddening residuals is also the region of lowest dust temperature.

A similar analysis with the S14 data yields residuals that are nearly flat with temperature within the uncertainties. However, at these low $E(B-V)$, the S14 map has much lower signal-to-noise than SFD or MF. Therefore, it is not possible to entirely rule out a systematic trend with dust temperature in the \ion{H}{i}-based reddening map based on this comparison.

\begin{figure}[tp]
	\includegraphics[width=\columnwidth]{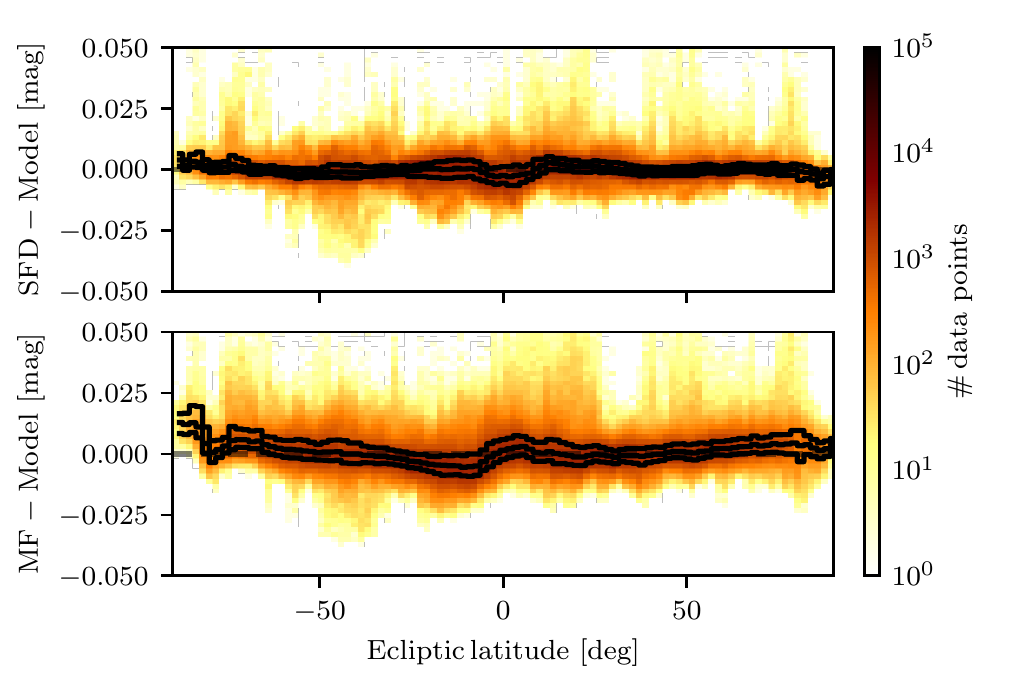}
	\caption{Differences of the present model and SFD ({\bf top}) and MF ({\bf bottom}) maps as a function of ecliptic latitude. The contour lines indicate 25, 50, and 75\% of the data points. While some trends are evident, it is difficult to disentangle possible contributions from the zodiacal light and residuals arising in regions at low ecliptic latitude which are also coincidentally close to the Galactic plane.}
	\label{fig:residuals_vs_ecliptic_latitude}
\end{figure}

\begin{figure}[tp]
	\includegraphics[width=\columnwidth]{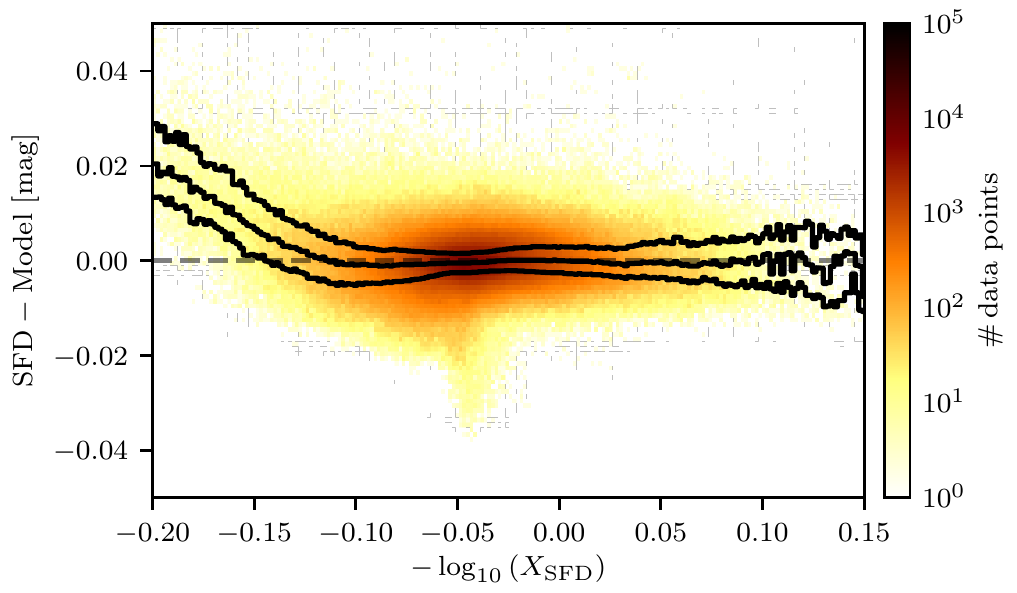}
	\caption{Residuals vs. SFD temperature corrections $X_{\rm SFD}$. The contour lines indicate 25, 50, and 75\% of the data points. While a strong trend is observed at high values of $X_{\rm SFD}$, it is unclear whether these residuals are due to faulty temperature corrections in the SFD map or to H$_2$ formation in cold gas.}
	\label{fig:residuals_vs_xmap}
\end{figure}

Another source of systematics affecting dust emission-based reddening maps is the contamination by zodiacal light. To investigate this, we plot our reddening residuals as a function of ecliptic latitude (Figure~\ref{fig:residuals_vs_ecliptic_latitude}). If dust in the Solar System were systematically affecting FIR-based reddening estimates, we would expect an increased scatter or a bias between dust-dependent and dust-independent reddening data at low ecliptic latitudes. We do observe some structure in this particular visualization of the residuals, but find our maps at low ecliptic latitude are dominated by a region of high residuals located coincidentally close to the Galactic plane. It is therefore difficult to definitively attribute the observed trends to systematics in the dust-based maps.

Given that little correlation is observed between the residuals with respect to SFD and the dust temperature correction or ecliptic latitude, and also that IR emission from galaxies has been demonstrated capable of inducing errors in the SFD reddenings at levels of $\sim 10$\,mmag \citep{Yahata+etal_2007}, we suggest that the observed discrepancies are due in large part to the misattribution of extragalactic IR emission to Galactic dust. If so, use of the SFD reddening map in cosmological surveys measuring, e.g., baryon acoustic oscillations could induce systematic biases in the recovered cosmological parameters. We defer investigation of the magnitude of this effect to future study but note that the \ion{H}{i}-based reddening map presented here is robust to this systematic.

\begin{figure}[tp]
	\includegraphics[width=\columnwidth]{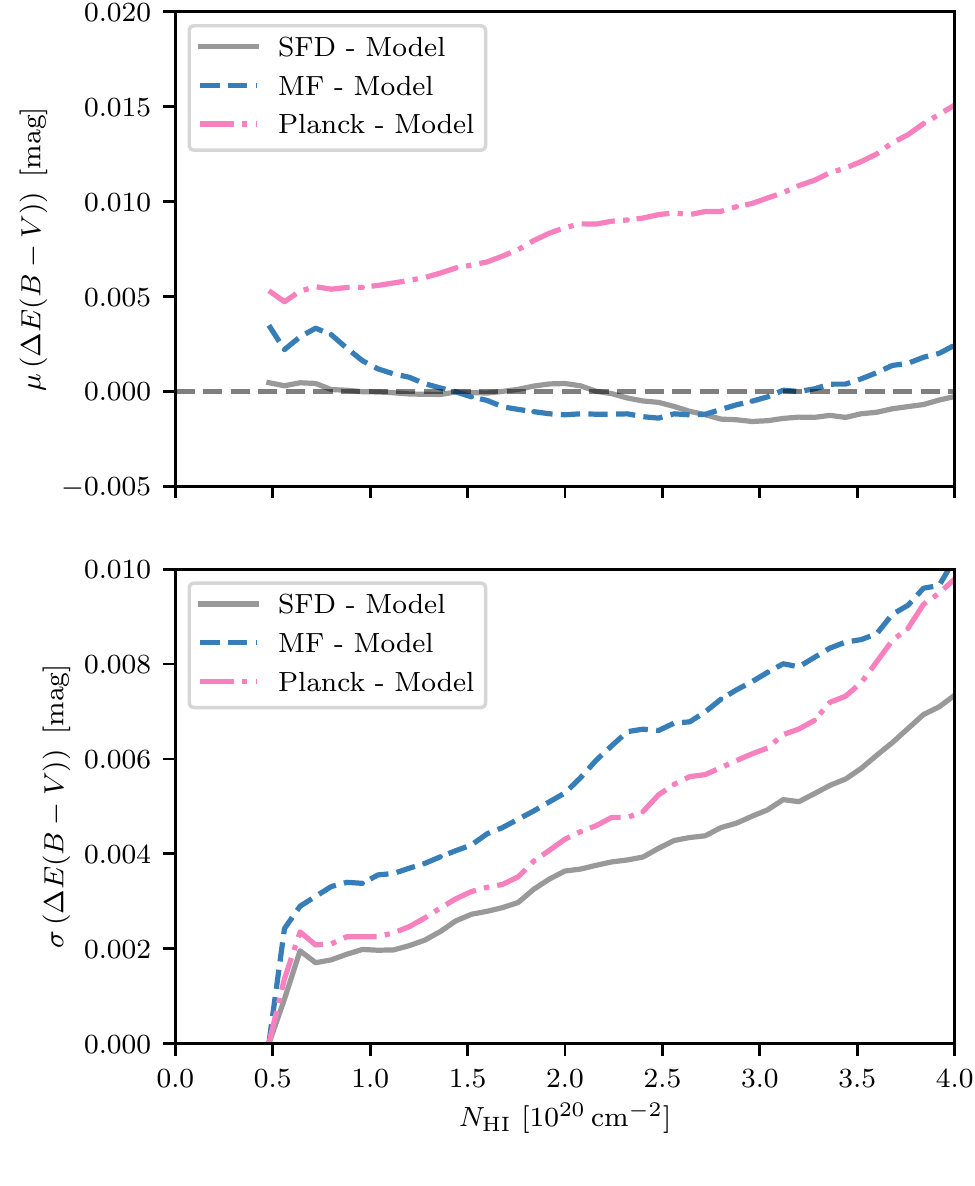}
	\caption{Median difference ({\bf top}) and standard deviation ({\bf bottom}) of residuals between the \ion{H}{i} based reddening map and dust emission based maps as a function of \ion{H}{i} column density.}
	\label{fig:residual_stddev_extern}
\end{figure}

Finally, we compare a number of dust-based reddening maps to our \ion{H}{i}-based map in Figure~\ref{fig:residual_stddev_extern}. In addition to the SFD and MF maps, we include the {\it Planck} $E(B-V)$ map, which is based on the dust radiance \citep{Planck_2013_XI}. In the top panel we compare the median residual with respect to the \ion{H}{i}-based map at fixed $N_\ion{H}{i}$. While the SFD map is flat to within a fraction of a mmag, the MF map varies by $\simeq 3$\,mmag over this range of $N_\ion{H}{i}$ and with a slight positive offset. In contrast, the {\it Planck} reddening map has an offset of $\simeq 5$\,mmag and varies by 10\,mmag over this range of column densities. Both S14 and MF noted systematic discrepancies between the {\it Planck} reddening map and the stellar data with increasing $E(B-V)$, which is consistent with the trends observed here though extending to much higher $E(B-V)$ than we consider.

In the bottom panel of Figure~\ref{fig:residual_stddev_extern}, we plot the standard deviation of the residuals at fixed $N_\ion{H}{i}$. With respect to all maps we observe increasing standard deviation with increasing $N_\ion{H}{i}$, indicative of H$_2$ formation inducing scatter in the \ion{H}{i}-based reddenings. The SFD map has the smallest scatter overall, though this is perhaps expected as it was employed to calibrate our model. The {\it Planck} map has slightly less scatter than the MF map over the full range, indicating that the bias noted in the top panel could potentially be corrected through recalibration, at least in the low column density regime.

\section{Discussion}
\label{sec:discussion}

\subsection{Model Extensions}

While \ion{H}{i} emission is an excellent tracer of dust at low column densities, the $\simeq 61$\% of the sky at higher column densities remains inaccessible due to the presence of H$_2$. We therefore explore the ability of tracers of dense gas to reliably extend our model to higher column densities.

\begin{figure}[tp]
	\includegraphics[width=\columnwidth]{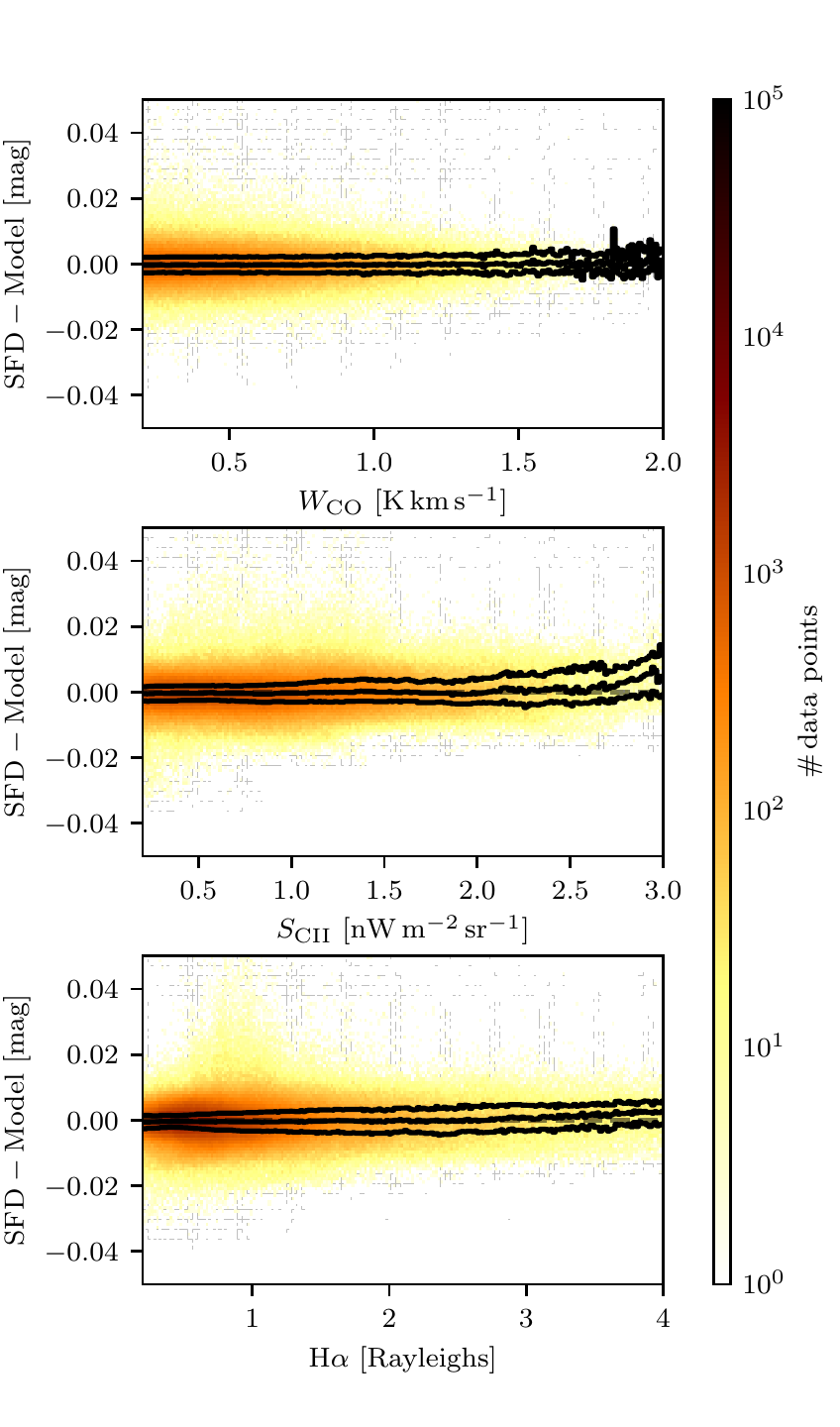}
	\caption{Residuals between the \ion{H}{i}-based reddening map and the SFD map as a function of different tracers of molecular and ionized gas. The \textbf{top} panel shows $W_{\rm CO}$ from \citet{Planck_2015_X}, the \textbf{middle} panel shows \ion{C}{ii} from DIRBE, and the \textbf{bottom} panel shows $\rm H\alpha$ intensity from \citet{Finkbeiner_2003}. The contour lines indicate 25, 50, and 75\% of the data points. Strong systematic trends are not seen for any tracer, although at high values of both \ion{C}{ii} and H$\alpha$ emission the reddening is slightly underpredicted by the \ion{H}{i}.}
	\label{fig:residuals_vs_ism_tracers}
\end{figure}

The first and most straightforward extension is use of CO as a tracer of molecular gas. In the top panel of Figure~\ref{fig:residuals_vs_ism_tracers} we plot our residuals with respect to the SFD map as a function of $W_{\rm CO}$ determined from full-sky fits to the {\it Planck} data \citep{Planck_2015_X}. However, we find no apparent correlation between our residuals and the CO emission. This is likely due to the sensitivity of the CO data, which is insufficient to detect the molecular cirrus present at high Galactic latitudes. A secondary effect is the presence of CO-dark molecular gas \citep[e.g.][]{Wolfire+etal_2010} which cannot be overcome simply with more sensitive CO maps.

Since \ion{C}{ii} forms at lower densities in the ISM than CO, it can be used to trace some of the CO-dark molecular gas \citep{Langer+etal_2010, Pineda+etal_2013, Pineda+etal_2017}. In the middle panel of Figure~\ref{fig:residuals_vs_ism_tracers}, we plot our residuals against \ion{C}{ii} emission measured by FIRAS\footnote{\url{https://lambda.gsfc.nasa.gov/product/cobe/firas\_products.cfm}} \citep{Bennett+etal_1994}. Despite the low resolution of the FIRAS data ($\sim 7\,\rm deg$) and the faintness of the signal, we do see a trend at very high \ion{C}{ii} intensities where the \ion{H}{i} underestimates the total reddening. More sensitive \ion{C}{ii} maps at higher angular resolution may therefore be able to significantly improve our reddening estimates and extend them to higher column densities.

High-energy tracers of total gas column density such as X-rays or $\gamma$-rays can also be used to get a handle on the CO-dark molecular gas and its associated reddening. However, the relation between this high-energy data and FIR data is not very well understood - the low resolution of the $\gamma$-ray data, changes in the dust radiative properties, as well as the phase transition from \ion{H}{i} to dark gas limit the analysis \citep{Planck_2015_XXVIII}.

Hence, we conclude that while additional tracers of the molecular ISM such as CO or \ion{C}{ii} might help to overcome the limitations at high reddenings \citep{Allen+etal_2015, Pineda+etal_2017}, the current observational data are insufficient to significantly improve our reddening estimates over the large scales and diffuse regions we consider in this work. It might prove insightful to perform a similar analysis over a limited sky area where one or more of these tracers have been measured with high sensitivity. 

Finally, we explored the possibility of dust-correlated ionized gas by employing the full-sky, 6' resolution $\mathrm{H}\alpha$ map by \citet{Finkbeiner_2003}\footnote{\url{https://lambda.gsfc.nasa.gov/product/foreground/fg\_halpha\_get.cfm}}. We plot the $\mathrm{H}\alpha$ emission against our reddening residuals in the bottom panel of Figure~\ref{fig:residuals_vs_ism_tracers}. We find a slight correlation at high H$\alpha$ intensities where \ion{H}{i} underestimates the total reddening, but the residuals are quite flat overall. This is perhaps unsurprising since ionized gas is expected to have a minor contribution to the total gas column.

In addition to ancillary gas tracers, future analysis could also employ more of the available \ion{H}{i} information. Full use of all \ion{H}{i} spectral channels instead of manually-chosen velocity bins could lead to refinements in modeling the velocity-dependence of $E(B-V)/N_{\ion{H}{i}}$ \citep[see][]{Lenz+etal_2016}. Additionally, the velocity {\it dispersion} of the neutral atomic hydrogen is a proxy for the kinetic temperature of the gas. A Gaussian decomposition of the \ion{H}{i} spectral data could identify correlations between reddening and gas temperature, as might be expected in dense, possibly molecular structures \citep{Haud_2010, Roehser+etal_2016}, and thereby improve our reddening estimates.

It would be ideal to employ both the dust emission and \ion{H}{i} data simultaneously, as suggested by \citet{Peek_2013}, to produce accurate reddening estimates. The \ion{H}{i} emission could enable calibration of the dust temperature corrections at low column densities, which could then be used with the dust emission data to produce reddening estimates at all column densities. Additionally, the \ion{H}{i} data would mitigate errors due to contamination from the CIB and the zodiacal light. 

\subsection{Comparison of dust and \ion{H}{i} emission as tracers of reddening}

In Section~\ref{sec:compare}, we presented an extensive comparison of our \ion{H}{i}-based reddening map to those based on dust emission. The availability of high-fidelity reddening maps constructed with orthogonal dust tracers enables quantitative assessment of the sensitivity of results to errors in the reddening correction. In addition, each map has advantages that may benefit certain applications.

We find that the \ion{H}{i}-based reddening estimates are limited by the presence of molecular and perhaps ionized gas along the line of sight. This restricts use of the \ion{H}{i}-based reddenings to low column density, low reddening sightlines. Even in this regime, there are Galactic structures with molecular gas, and so the reddening errors can be correlated with these structures as noted in Section~\ref{sec:reddening_map}. As we demonstrate in Section~\ref{sec:compare}, the \ion{H}{i}-based reddenings are superior to the dust-based maps in this low column density regime based on comparisons to direct reddening determinations in the optical.

In contrast to the \ion{H}{i}-based reddening maps, those based on dust emission can probe the full range of column densities. Their principal limitations are applying a temperature correction to infer the dust column and contamination from dust that is not interstellar, notably the zodiacal light and the CIB. Both of these uncertainties have been shown to be small, typically of order mmag but up to tens of mmag \citep{Peek+Graves_2010}. However, as noted by \citet{Yahata+etal_2007}, reddening errors that are correlated with the CIB can be particularly problematic for studies undertaking precision probes of large scale structure. For these applications, the \ion{H}{i}-based reddenings may be preferred.

\section{Conclusions}
\label{sec:conclusions}

The principal conclusions of this work are as follows:

\begin{enumerate}
	\item For $N_{\ion{H}{i}} < 4\times 10^{20}\,\rm cm^{-2}$, \ion{H}{i} is an excellent predictor of dust reddening, with a mean value of $N_\ion{H}{i}/E(B-V) = 8.8\times 10^{20}\,\rm cm^{-2}\,mag^{-1}$. The linear relationship between $N_\ion{H}{i}$ and $E(B-V)$ breaks down at roughly this threshold, presumably due to the presence of H$_2$ gas.
   \item We find little to no reddening associated with gas with $|v_{\rm LSR}| > 90\,\rm km\,s^{-1}$, consistent with previous studies of high-velocity gas. 
    \item We have employed \ion{H}{i} emission data to create an $E(B-V)$ map covering the 39\% of the sky with $N_{\ion{H}{i}} < 4\times 10^{20}\,\rm cm^{-2}$, mostly at high Galactic latitude. This map is in excellent agreement with SFD, having a typical scatter (instrumental + systematic) of 5.1\,mmag. We provide the map at 16'.1 resolution gridded to a HealPix N$_{\rm side}$ = 1024.
    \item We confirm the systematic reddening errors in SFD identified by \citet{Peek+Graves_2010} using extragalactic observations. We argue that, in the low column density regime, the $E(B-V)$ map derived in this work is more reliable than those based on dust emission.
  \item The $E(B-V)$ map is made available online\footnote{\url{http://dx.doi.org/10.7910/DVN/AFJNWJ}} in HEALPix format, with an $\rm N_{side}$ of 1024 and an angular resolution of $16\overset{'}{.}1$. We also provide tools to measure the $E(B-V)$ for a particular sightline and to convert the observed reddening to extinction in various filters.
\end{enumerate}

\acknowledgments
{We thank Aaron Meisner, Josh Peek, Eddie Schlafly, and Benjamin Winkel for assistance with various data products and helpful conversations; Lars Fl\"oer, J\"urgen Kerp,  Guilaine Lagache, and Gina Panopoulou for valuable feedback; and the anonymous referee for comments which improved the manuscript. This research was carried out at the Jet Propulsion Laboratory, California Institute of Technology, under a contract with the National Aeronautics and Space Administration. EBHIS is based on observations with the 100-m telescope of the MPIfR (Max-Planck-Institut f\"{u}r Radioastronomie) at Effelsberg. The Parkes Radio Telescope is part of the Australia Telescope which is funded by the Commonwealth of Australia for operation as a National Facility managed by CSIRO. We acknowledge use of the Legacy Archive for Microwave Background Data Analysis (LAMBDA). Support for LAMBDA is provided by the NASA Office of Space Science. This research has made use of NASA's Astrophysics Data System, matplotlib \citep{Hunter_2007}, SciPy \citep{jones_scipy_2001}, NumPy \citep{van2011numpy}, as well as  Astropy, a community-developed core Python package for Astronomy \citep{2013A&A...558A..33A}. Some of the results in this paper have been derived using the HEALPix \citep{Gorski+etal_2005} package.

\bibliographystyle{yahapj}
\bibliography{references}

\end{document}